\documentclass[aps,prb,twocolumn,groupedaddress,floatfix,superscriptaddress, longbibliography]{revtex4-1}

\usepackage{amsmath}
\usepackage{amssymb}
\usepackage{graphicx}
\usepackage{mathrsfs}
\usepackage{float}
\usepackage[english]{babel}
\usepackage{dsfont}
\usepackage{epstopdf}
\usepackage{soul}
\usepackage{color}
\usepackage{braket}
\usepackage[colorlinks=true,
            linkcolor=magenta,
            urlcolor=blue,
            citecolor=blue]{hyperref}
\usepackage{appendix}          
\usepackage[dvipsnames]{xcolor}

\usepackage[normalem]{ulem}
\def\be{\begin{equation}}
\def\ee{\end{equation}} 
\def\bea{\begin{eqnarray}}
\def\eea{\end{eqnarray}} 
\def\ba{\begin{array}} 
\def\ea{\end{array}}

\def\rm{\mathrm}

\newcommand{\mv}[1]{\langle #1\rangle}

\newcommand{\unit}{1\!\!1}
\usepackage{comment}
\usepackage{cancel}

\emergencystretch=\maxdimen
\begin{document}

\title{Emergent chiral Higgs mode in $\pi$-flux frustrated lattices}

\author{M. Lanaro}
\affiliation{Dipartimento di Fisica e Astronomia ``G. Galilei", Università degli Studi di Padova, I-35131, Padova, Italy}
\affiliation{INFN Istituto Nazionale di Fisica Nucleare, Sezione di Padova, I-35131, Padova, Italy}
\author{L. Maffi}
\affiliation{Dipartimento di Fisica e Astronomia ``G. Galilei", Università degli Studi di Padova, I-35131, Padova, Italy}
\affiliation{INFN Istituto Nazionale di Fisica Nucleare, Sezione di Padova, I-35131, Padova, Italy}
\author{M. Di Liberto}
\affiliation{Dipartimento di Fisica e Astronomia ``G. Galilei", Università degli Studi di Padova, I-35131, Padova, Italy}
\affiliation{INFN Istituto Nazionale di Fisica Nucleare, Sezione di Padova, I-35131, Padova, Italy}
\affiliation{Padua Quantum Technologies Research Center, via Marzolo 8, I-35131 Padova, Italy}

\begin{abstract}

Neutral-atom quantum simulators provide a powerful platform for realizing strongly correlated phases, enabling access to dynamical signatures of quasiparticles and symmetry breaking processes. Motivated by recent observations of quantum phases in flux-frustrated ladders with non-vanishing ground-state currents, we investigate interacting bosons on the dimerized Benalcazar-Bernevig-Hughes (BBH) lattice in two dimensions—originally introduced in the context of higher-order topology. 
After mapping out the phase diagram, which includes vortex superfluid (V-SF), vortex Mott insulator (V-MI), and featureless Mott insulator (MI) phases, we focus on the integer filling case.
There, the MI/V-SF transition breaks the $\mathbb Z_2^{T}$ and U(1) symmetries, where $\mathbb Z_2^{T}$ corresponds to time-reversal symmetry (TRS). 
Using a slave boson description, we resolve the excitation spectrum across the transition and uncover a chiral Higgs mode whose mass softens at criticality, providing a dynamical hallmark of emergent chirality that manifests via the oscillation of the loop currents in quench dynamics. 
Our results establish an experimentally realistic setting for probing unconventional TRS-broken phases and quasiparticles with intrinsic chirality in strongly interacting quantum matter.

\end{abstract}
\maketitle

\section{Introduction}
Neutral atoms quantum simulators offer a versatile playground to study strongly correlated phenomena \cite{Nori2014,gross2017quantum}.
Beyond identifying equilibrium properties, monitoring real-time dynamics has given access to their collective excitations and quantum correlations \cite{gross2013,bernien2017, gross2019,prichard2025magnon,chen2025}.
A prominent role in symmetry-broken phases is played by amplitude (Higgs) excitations of the order parameter \cite{pekker2015amplitude}, observed, for example, in interacting bosonic \cite{bissbort2011detecting, endres2012higgs} and fermionic systems \cite{behrle2018higgs,Kell2024}, long-range cavity systems \cite{leonard2017monitoring}, Rydberg spin systems \cite{manovitz2025quantum} and predicted for dipolar supersolids \cite{Pfau2019}.
Identifying unconventional quasiparticles in strongly correlated quantum systems, including those appearing in lattice gauge theories \cite{aidelsburger2022cold,halimeh2025cold}, is currently a major goal of quantum simulation platforms.

\begin{figure}[!ht]
    \centering    \includegraphics[width=0.98\linewidth]{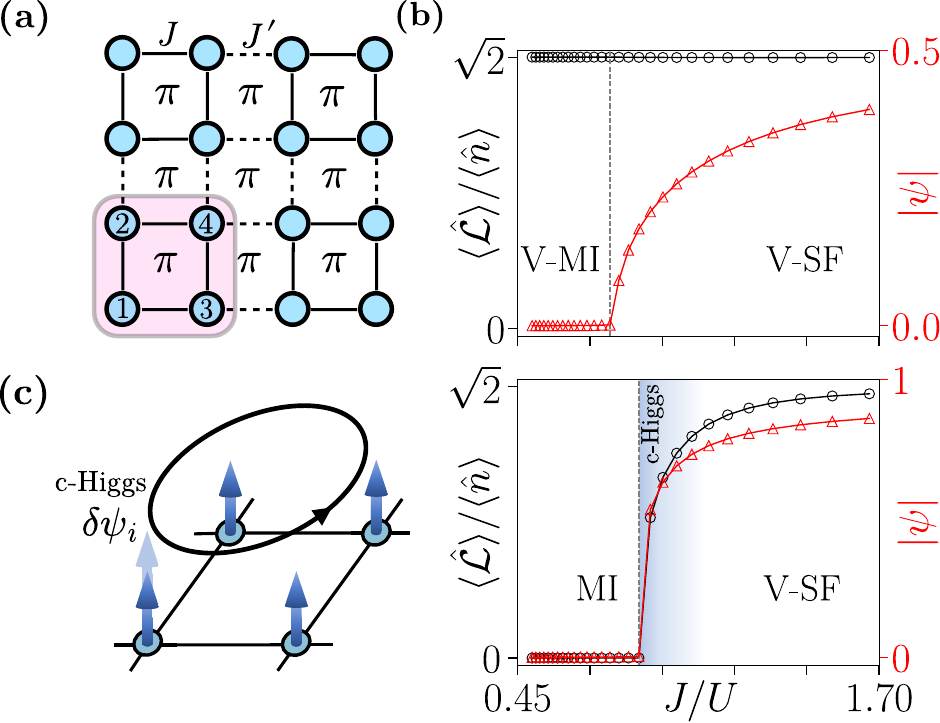}
    \caption{Model and main results. \textbf{(a)} Schematics of the BBH lattice with dimerized hopping $J$, $J'$ and uniform $\pi$ flux for interacting bosons. The unit cell is highlighted. \textbf{(b)}  Ground-state loop current $\mv{\hat{\mathcal{L}}}$ (black circles) and superfluid order parameter $|\psi|$ (red triangles) within cluster Gutzwiller approximation marking phase transitions between vortex superfluid (V-SF) and vortex Mott insulating (V-MI) phases for filling $\nu = 1/4$ (top) and between V-SF and Mott-insulating (MI) phases for $\nu = 1$ (bottom) for $J'/J=0.03$. 
    The region displaying a chiral Higgs excitation is highlighted.
    \textbf{(c)} Pictorial representation of chiral Higgs mode dynamics within a plaquette, where $\delta \psi_{i}$ indicates the amplitude excitation and the mode exhibits a finite loop current.}
    \label{Fig1}
\end{figure}

Synthetic gauge fields \cite{jaksch2003creation, aidelsburger2011, goldman2016topological} provide a key ingredient to simulate correlated phases that break time-reversal symmetry (TRS) as for Fractional Quantum Hall states \cite{cooper2019topological, laughlin1983anomalous,prange1987quantum}.
These states can host quasiparticle excitations that include Abelian and non-Abelian anyons \cite{halperin1984statistics, stern2008anyons} as well as chiral graviton modes \cite{haldane2011geometrical, liang2024evidence}, thus stimulating efforts for their identification in synthetic matter systems \cite{bacciconi2025theory, xavier2025chiral}.
Recent progress in this direction has been demonstrated through the realization of Laughlin and Pfaffian states with few atoms \cite{leonard2023realization,lunt2024realization, kwan2026pfaffian}.
To date, many-body topologically ordered phases however remain still a challenge for analog quantum simulators, as compared to digital realizations \cite{evered2025probing}.
However, TRS broken phases without topological order can also display strongly-correlated properties manifesting via unconventional quasiparticles, which is the focus of this work.

Recent experimental advances have demonstrated that many-body quantum phases that break TRS via specific ground-state current patterns and correlations are within reach. For example, quasi-1D optical lattice geometries with fermions in synthetic dimensions \cite{zhou2023observation} have shown the signature of a Hall response~\cite{greschner2019}. Bosonic ladder experiments \cite{atala2014observation, impertro2025} have demonstrated the preparation of a Meissner phase \cite{orignac2001}, a strongly interacting gapped phase with broken TRS.  
Such phase displays leg-currents that can be viewed as a precursor of FQH state edge modes in the sense of the wire construction \cite{kane2002, Mazza2017}. 
In this context, correlated bosonic quantum phases with ground-state currents or vortices offer an intriguing quantum simulation scenario. 
The interplay of interaction with magnetic and geometrical frustration can indeed give rise to a variety of quantum phases with broken TRS whose excitation properties remain largely unexplored.  
These include the Meissner phase mentioned above, or phases characterized by loop currents forming  staggered, commensurate or incommensurate vortex arrays~\cite{Polini2005,MoraisSmith2008, cha2011, Dhar2012, greschner2013, altman2014, Kolley_2015, piraud2015, greschner2015, petrescu2017, greschner2016, halati2023, DiLibertoGoldman, barbiero2023, dasgupta2025chiral,hasan2025superfluid}, which can display distinct Hall signatures \cite{Citro2025}. 
The recent experimental developments \cite{impertro2025} suggest some of them to be within reach. 
In 2D, this interplay can also give rise to topological bosonic insulating phases~\cite{sorensen2005,  hafezi2007,pollman2015,boesl2022,Yao2024} characterized by quantized Hall responses and topological order \cite{wen2017}, which remain challenging to realize, as we discussed above.

In this work, we examine  ground state and low-lying excitations of interacting bosons on a flux-frustrated dimerized model in two dimensions (Fig.~\ref{Fig1}(a)),
relevant to higher-order topological insulator models \cite{BBH_Science, BBH_PRB} and recent optical lattice experiments \cite{impertro2025}.  
We identify vortex superfluid (V-SF), vortex Mott insulator (V-MI), and featureless Mott insulator (MI) phases via DMRG numerical simulations and a cluster Gutzwiller approach, Fig.~\ref{Fig1}(b).
The MI/V-SF transition at integer filling spontaneously breaks the $\mathbb Z_2^T$ and U(1) symmetries, where $\mathbb Z_2^T$ indicates TRS. 
A similar mechanism has been suggested to occur in the context of flux-frustrated 2D Josephson junctions arrays \cite{Polini2005} and identified, for example, in $p$-band models \cite{li2011} and twisted superfluids \cite{Sengstock2012, Jurgensen2015}.  
In our 2D model, the key features of this phase transition are reflected in the excitation spectrum, where we identify a chiral Higgs mode carrying nonvanishing vorticity, Fig.~\ref{Fig1}(c). 
This mode softens across the transition and can be revealed by quench dynamics manifesting in observable oscillations of the currents.
To the best of our knowledge, this type of quasiparticle has not been identified before in bosonic models.
Our results based on a minimal and experimentally accessible construction provide a pathway for investigating unexplored properties of strongly-correlated quantum matter, as for the emergence of unconventional quasiparticles with intrinsic chirality.

The paper is organized as follows. In Sec.~\ref{sec:II}, we introduce the model, the cluster mean-field approximation and discuss the corresponding phase diagram in 2D; in Sec.~\ref{sec:III}, we perform a DMRG analysis of the model in a ladder geometry; in Sec.~\ref{sec:IV}, we employ a slave-boson method to compute the excitations in the superfluid phase near the critical point; in Sec.~\ref{sec:V}, we discuss the results and draw the conclusions.

\section{Model and quantum phases}
\label{sec:II}
We consider soft-core bosons on a dimerized lattice depicted in Fig.~\ref{Fig1}(a) with uniform $\pi$-flux per plaquette, and thus corresponding to the Benalcazar-Bernevig-Hughes (BBH) geometry \cite{BBH_Science, BBH_PRB}.
In this work, we study the strongly dimerized limit $J\gg J'$, where $J$ and $J'$ describe the intra- and the inter-plaquette coupling amplitudes, respectively. 
The free particle Hamiltonian is thus
\begin{equation}
\hat{\mathcal H}_0 = \hat{\mathcal H}_J + \hat{\mathcal H}_{J'}\,.
\end{equation}
Onsite interactions are described by   
\begin{equation}
\mathcal{\hat{H}}_{\text{onsite}} = \frac{U}{2} \sum_{p}\sum_{i=1}^{4} \hat{n}_{i,p}^{} (\hat{n}_{i,p}^{} -1 ) - \mu \sum_{p}\sum_{i=1}^4\hat{n}^{}_{i,p}\,,
\end{equation}
where the number operator $\hat{n}^{}_{i,p}=\hat{b}^\dagger_{i,p}\hat{b}^{}_{i,p}$ acts on the sublattice site $i$ of the plaquette $p$ according to the convention in Fig.~\ref{Fig1}(a), $U>0$ is the interaction strength and $\mu$ the chemical potential. 
The model $\hat{\mathcal{H}}=\hat{\mathcal{H}}_0^{} + \hat{\mathcal{H}}^{}_{\text{onsite}}$ describes a BBH Bose-Hubbard model and can be readily realized in optical superlattices via synthetic magnetic fields \cite{impertro2025}.

\begin{figure}
    \centering    \includegraphics[width=0.88\linewidth]{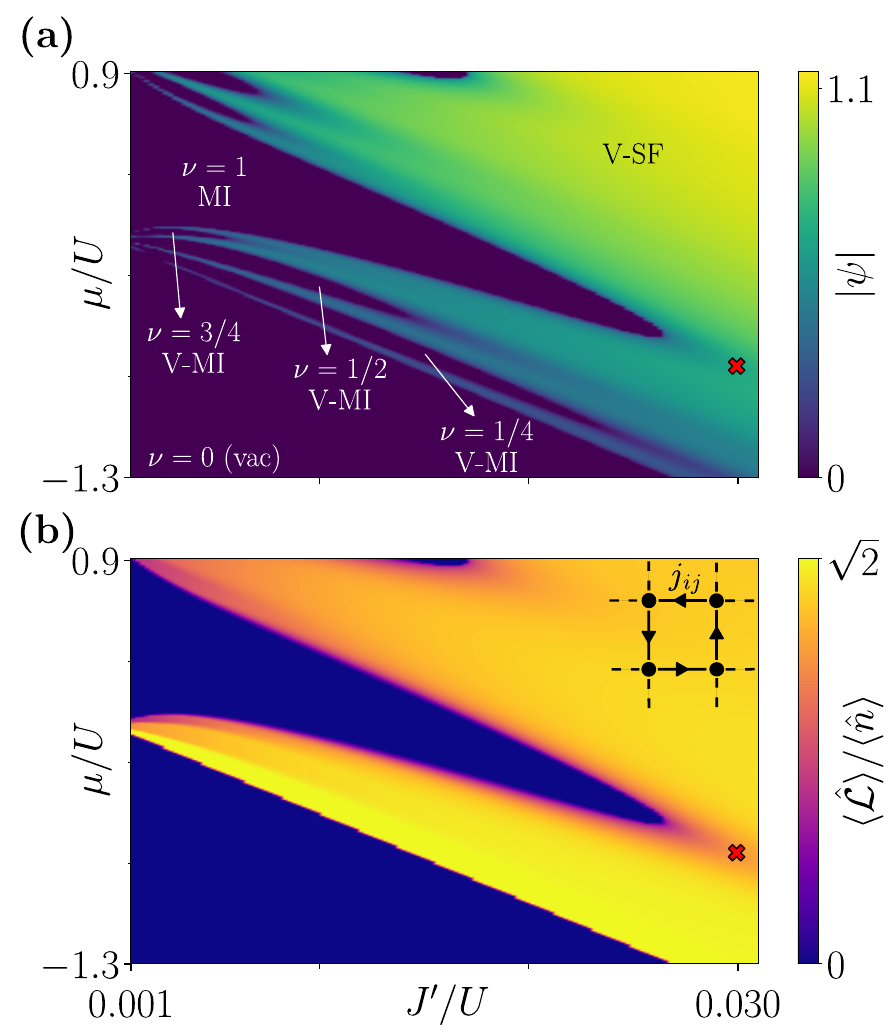}
        \caption{Phase diagram for $J'/J = 0.03$ of the 2D BBH model via cluster Gutzwiller approximation ($n_{\text{max}}=5$). \textbf{(a)} Colormap of the superfluid order parameter $|\psi|$ showing Mott phase (MI), vortex Mott phases (V-MI) at different fillings and the vortex superfluid phase (V-SF). \textbf{(b)} Corresponding loop current per particle across the phase diagram. Notice that only the Mott lobe at $\nu=1$ has zero loop current, while fractional fillings exhibit $\langle \hat{\mathcal{L}} \rangle \neq 0$, indicating a V-MI phase. Inset shows the current pattern in V-SF and V-MI phases.}
    \label{Fig2}
\end{figure}

In the strongly dimerized limit, each plaquette plays the role of a super-site. 
Here we perform our analysis using a cluster Gutzwiller ansatz (cGWA) \cite{Luhmann_GW, yamamoto2012, huerga2013, yamamoto2014, paul2025interacting}, generalizing the more conventional single-site Gutzwiller ansatz \cite{krauth1992, jaksch1998}, allowing us to capture quantum phases at fractional filling as well as to retain quantum correlations within a plaquette~\footnote{A single-site ansatz is described by a wavefunction $|\psi_i\rangle = \sum_n c_n^{} |n\rangle$. Using such an ansatz, the fractional filling condition, \emph{e.g.} $\mv{\hat n^{}_i}=1/2$, necessarily implies $\mv{\hat b_i^{}}\neq0$. This forbids the possibility to capture phase transitions into insulating phases with fractional filling within a single-site Gutzwiller ansatz.}. 
Specifically, we perform a decoupling by considering a plaquette product state of the form
\be
|\Phi \rangle_{\text{cl}} \equiv \bigotimes_{p} \biggl( \sum_{n^{}_{1}n^{}_{2}n^{}_{3}n^{}_{4}} c^{(p)}_{n^{}_{1}n^{}_{2}n^{}_{3}n^{}_{4}}|n^{}_{1}n^{}_{2}n^{}_{3}n^{}_{4}\rangle_{p}^{} \biggr),
\ee
where $n_i$ denotes the Fock occupation number on the sublattice site $i$.
Within this cluster mean-field decoupling, the Hamiltonian reads
\begin{equation}
\hat{\mathcal{H}} \approx \hat{\mathcal{H}}_{\text{cl}}^{} - \sum_{\substack{\langle p,q \rangle \\ \langle i,j \rangle }} \bigl[J'_{ij} \bigl(\hat{b}_{i,p}^\dagger \psi^{j,q}_{} -(\psi^{j,q})^*\psi^{i,p} \bigr) +\text{h.c.} \bigr]\,,
\end{equation}
where $\psi^{i,p}_{} \equiv \langle \hat{b}_{i,p}^{} \rangle$ is the superfluid order parameter and $\hat{\mathcal H}_\text{cl}^{}$ is the cluster Hamiltonian obtained by considering terms in $\hat{\mathcal H}$ that contain only operators acting on a single plaquette~\cite{SuppMat}. 
This allows us to write $\hat{\mathcal H}\approx\sum_p \hat{\mathcal H}_\text{MF}^{(p)}$, where $\hat{\mathcal H}_\text{MF}^{(p)}$ contains only operators acting on a single plaquette $p$. 
We solve for the ground-state by assuming (plaquette) translational symmetry for a two-dimensional (2D) lattice, namely $\psi^{i,p} = \psi^{i}$ and by diagonalizing $\hat{\mathcal H}_\text{MF}^{(p)}$ to find the cluster wavefunction parameters $c^{(p)}_{n_{1}^{}n_{2}^{}n_{3}^{}n_{4}^{}}$ via a self-consistent algorithm (truncating $0 \leq n_i < n_\text{max}$ corresponding to a plaquette Hilbert space of dimension $d=n_\text{max}^4$).
The resulting phase diagram is shown in Fig.~\ref{Fig2}(a). 

In the strongly dimerized case, Ref.~\cite{DiLibertoGoldman} showed that at low energies each plaquette $p$ encodes two degenerate orbitals, $|\pm\rangle_p$, that are eigenstates of an emergent (plaquette) angular momentum operator $\hat L_z^{(p)}$ commuting with $\hat{\mathcal H}$, namely $\hat L_z^{(p)}|\pm\rangle_p=\pm|\pm\rangle_p$, and carry (opposite) non-vanishing loop currents. 
At the plaquette level, interactions take the approximate form $\mathcal{\hat{H}}_{\text{onsite}}\approx -(U/16)\sum_p [\hat L_z^{(p)}]^2 + \dots$ and favor the formation of ground-state currents, 
\begin{equation}
\hat{j}_{ij} = i  \big(J_{ij}\,\hat{b}_{i}^{\dagger}\hat{b}_{j}^{} - \text{h.c.} \bigr)\,,
\end{equation}
maximizing the angular momentum and thus leading to spontaneous breaking of TRS. Indeed, the loop current operator within a plaquette, $\hat{\mathcal{L}}^{(p)} = J^{-1}\sum_{\Box} \hat{j}_{ij}\approx \sqrt 2 \hat L_z^{(p)}$ takes a nonzero expectation value, where $\sum_\Box$ indicates the oriented sum on the plaquette $p$.
In the weakly-interacting limit, this corresponds to a macroscopic occupation of either $|+\rangle$ or $|-\rangle$, similarly to what happens in higher-band systems \cite{girvin2005, Liu2006, wirth2011evidence, li2016physics, Li_TRS}.
However, this feature persists in the strongly-correlated superfluid phase. The value of $\mv{\hat{\mathcal{L}}}$ across the phase diagram is shown in Fig.~\ref{Fig2}(b) and the corresponding phase is a V-SF that breaks $\mathbb Z_2^T \times$U(1). 
Strong interactions renormalize the loop current, such that its value per particle is no longer exactly quantized in units of $\sqrt 2$ as in the low-energy theory.

When interactions dominate, $U\gg J'$, different quantum phase transitions occur depending on the filling $\nu$ (bosons per site).
At filling $\nu=1/2$, we confirm the transition identified via DMRG in Ref.~\cite{DiLibertoGoldman} where the system enters a vortex Mott insulating phase (V-MI), namely an incompressible translational invariant phase lacking phase coherence and with loop currents \cite{piraud2015,impertro2025} localized at the plaquettes, see inset of Fig.~\ref{Fig2}(b).  
At filling $\nu=1/4$, namely one particle per plaquette, we also find a V-MI phase. 
This is in contrast with what happens in $p$-band models, where the insulating phase corresponds to a staggered paramagnet with no orbital order \cite{girvin2005, li2011, li2016physics}. 
To understand why the quantum fluctuations in our model still favour a V-MI phase we perform second order perturbation theory at $J'/U \ll 1$ for filling $\nu=1/4$. 
For this, we consider the chiral basis introduced above, $|\pm\rangle$, which represents the two chiral single-particle modes of a $\pi$-flux plaquette and can be mapped onto an effective spin-1/2 degree of freedom.
Perturbative calculations yield the quantum XXZ 2D Hamiltonian 
\begin{equation}
\hat H_{\text{eff}}= - \frac{5}{2}\frac{J'^{2}}{U} -\frac{J'^{2}}{4U}\sum_{\mv{p,q}} \Bigl[
3\hat{\sigma}^{z}_p\hat{\sigma}^{z}_q
+ \hat{\sigma}^{x}_p\hat{\sigma}^{x}_q
+ \hat{\sigma}^{y}_p\hat{\sigma}^{y}_q \Bigr]\,.
\end{equation}
Differently from $p$-band systems, where $\hat H_{\text{eff}}$ is classical at leading order, we obtain a quantum effective model. 
Furthermore, the parameters in $\hat H_{\text{eff}}$ set the system in the easy-axis regime \cite{Loh1985} and put the ground-state in the ferromagnetic phase \cite{Cuccoli2001, peter2012}, which corresponds to a V-MI phase.
Finally, we also identified a transition to an insulating phase at $\nu=3/4$ in a small regime of phase space.

At integer filling, $\nu=1$, a completely different scenario takes place, which is the core of this work. 
Correlations within a plaquette will be suppressed when $U\gg J,J'$ and the system displays a transition onto a Mott insulating (MI) phase with a unique ground-state, asymptotically connected to the product state $|1111\dots\rangle$.
Unlike the previous cases, a distinct symmetry breaking pattern thus takes place. 
In addition to U(1) (equivalently O(2)), the superfluid ground-state also displays different time-reversal properties as compared to the nearby Mott phase, which therefore correspond to the breaking of $\mathbb{Z}_2^T\times$U(1), see Fig.~\ref{Fig1}.
This fact will have dramatic consequences in the excitation spectrum, as we will discuss down below.

\begin{figure}[t]
    \centering
\includegraphics[width=0.98\linewidth]{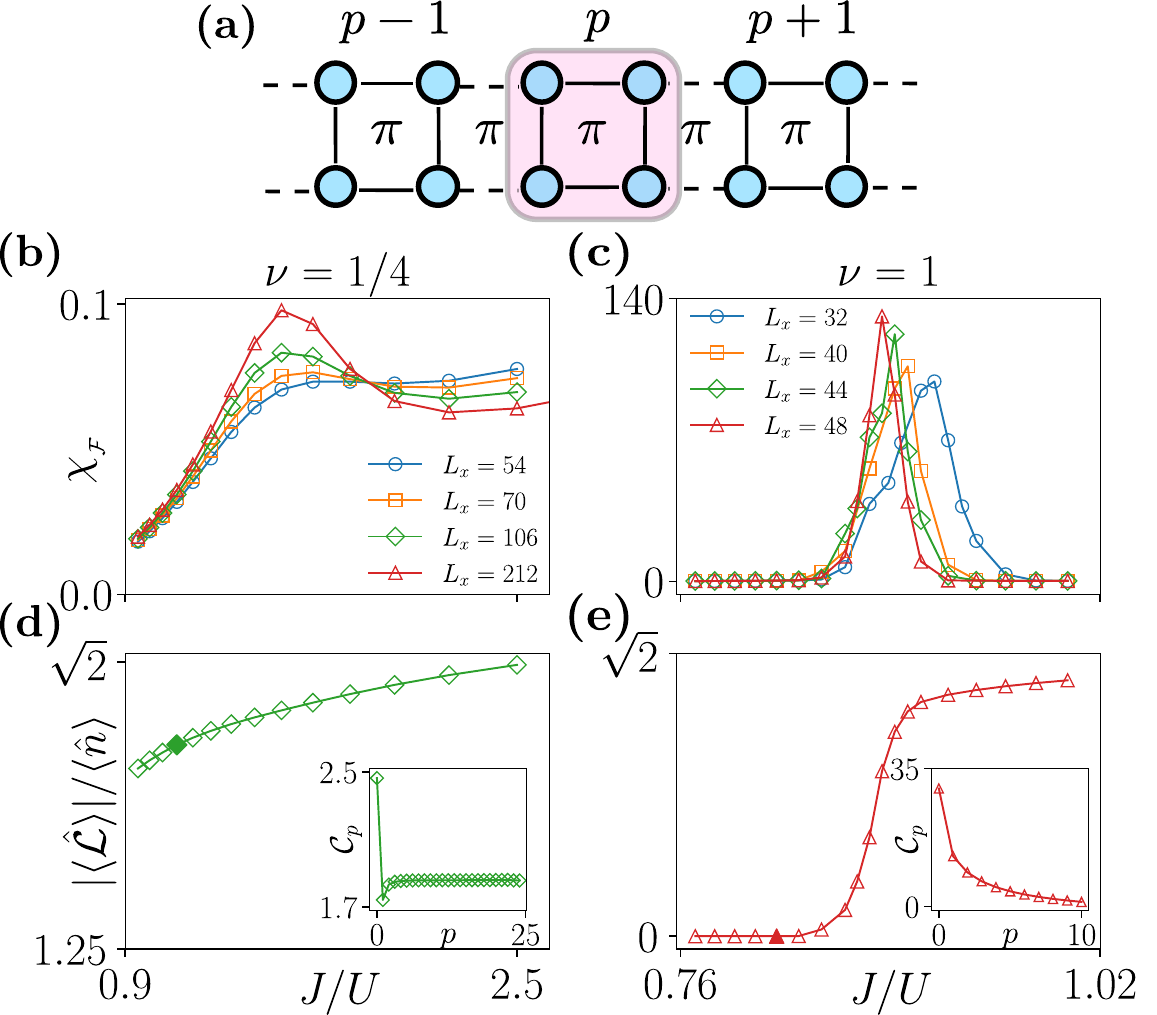}
    \caption{DMRG numerical results for a ladder with $J'/J=0.1$. \textbf{(a)} Ladder geometry for DMRG simulations. \textbf{(b)}-\textbf{(c)} Fidelity susceptibilities $\chi_{\mathcal{F}}$ for $\nu=1/4$ and $\nu=1$. The peaks growing with system size are indication of the phase transition. \textbf{(d)}-\textbf{(e)} Loop current per particle $|\langle \hat{\mathcal{L}} \rangle | / \langle \hat{n} \rangle $ for $L_{x}=106$ ($\nu=1/4$) and for $L_{x}=48$ ($\nu=1$). The $\nu=1$ curve and the drop of loop current show the V-SF / MI phase transition. Insets show the $\mathcal C_p$ correlator corresponding to the filled symbols.
    } 
    \label{Fig3}
\end{figure}

\section{DMRG analysis on a ladder}
\label{sec:III}
We complement our findings by performing DMRG simulations via the iTensor library \cite{DMRG} on a ladder of dimensions $L=2\times L_x$ shown in Fig.~\ref{Fig3}(a) for $J'=0.1J$.
We identify the transitions by computing the fidelity susceptibility, 
\begin{equation}
\chi_{\mathcal{F}} = - \frac{\langle \psi ( U+\delta U) | \psi ( U)\rangle - 2 + \langle \psi ( U-\delta U) | \psi ( U)\rangle}{\delta U^{2}L}\,,
\end{equation}
expected to diverge at the transition point \cite{carrasquilla2013}.
We find two distinct peaks displayed in Fig.~\ref{Fig3}(b),(c): one occurs at $U/J \sim 0.65$ for $\nu=1/4$ and another one at $U/J \sim 1.13$ for $\nu=1$, whereas the transition at $\nu=1/2$ was identified at $U/J\sim 1.2$ in Ref.~\cite{DiLibertoGoldman}. 
The plaquette loop current $\mv{\hat{\mathcal{L}}}$ manifests distinct behaviour for the two types of transitions, see Fig.~\ref{Fig3}(d),(e). 
While for $\nu=1/4$ the loop current remains finite and smooth across the critical point, for $\nu=1$ we observe a clear drop signaling that the ground-state restores time-reversal symmetry in the insulating phase 
\footnote{A similar behaviour is identified also in a small 2D system via Tree Tensor Network calculation, as shown in Ref.~\cite{SuppMat}.}.
In the insets, we estimate the loop current correlations between plaquettes, $\mathcal{C}_p = \langle\mathcal{\hat{L}}_{0} \mathcal{\hat{L}}_{p}\rangle$, showing that the $\nu=1/4$ insulating phase displays ferromagnetic long-range order as predicted by the perturbative analysis discussed above; instead, for $\nu=1$, $\mathcal{C}_p$ decays exponentially as expected in a featureless MI phase.
The identification of the transition at filling $\nu=3/4$ with DMRG has instead proven to be computationally challenging, as already anticipated by the narrow parameters range predicted by the cGWA phase diagram.

\begin{figure}[!b]
    \centering
    \includegraphics[width=1
    \linewidth]{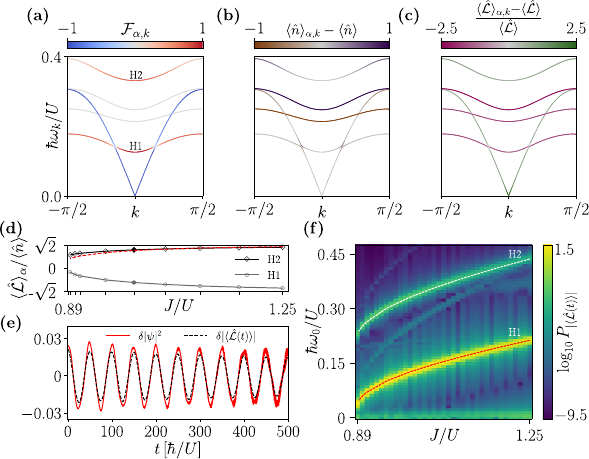}
    \caption{Low-energy excitations in the V-SF phase for filling $\nu = 1$. 
    \textbf{(a)}-\textbf{(c)} Quasiparticle excitations for \mbox{$k_x = k_y = k$, $J'/U = 0.03$}, $J/U = 1.0$, and $\mu/U = -0.68$, corresponding to the cross symbols in Fig.~\ref{Fig2}. 
    Spectra are colored based on the values of \textbf{(a)} flatness $\mathcal{F}_{\alpha, k}$,  \textbf{(b)} plaquette density and \textbf{(c)} loop current $\langle \mathcal{\hat{L}} \rangle_{\alpha,k}$. The index $\alpha=1,\dots, 5$ labels the modes for increasing value of energy
\textbf{(d)} Loop current per particle, $\langle \hat{\mathcal{L}} \rangle_{\alpha} / \langle \hat{n} \rangle$, for H1, H2 and superfluid ground state (red dashed line). Filled symbols correspond to spectra  shown in the upper panel.
\textbf{(e)} cGWA time evolution of loop current (black dashed line) and local order parameter (red solid line) oscillations normalized to their ground-state values following a quench of a staggered flux $\pm\delta\theta$ per plaquette at unit filling. Parameters as in (a)-(c) and $\delta\theta = 0.01 (J-\bar{J})/\bar{J}$ with $\bar{J}=0.88\,U$.
\textbf{(f)} Power spectrum of the loop current time evolution for different values $J/U$. Dashed lines highlight the frequencies corresponding to H1 (red) and H2 (white) modes.
}
    \label{Fig4}
\end{figure}

\section{Excitations near the critical point}
\label{sec:IV}
Quantum phase transitions  with different symmetry-breaking mechanisms are expected to display low-energy excitations with distinct signatures \cite{sachdev2023quantum}.
The fractional filling  transitions only break U(1) symmetry and thus display analogous properties as the conventional Mott-superfluid transitions \cite{altman2002, pekker2015amplitude, sachdev2023quantum}.  
In this case, besides a low-energy gapless Goldstone mode, an amplitude (Higgs) excitation softens, as it was experimentally confirmed for the standard Bose-Hubbard model \cite{bissbort2011detecting, endres2012higgs}. 
At integer filling, the ground-state additionally breaks TRS across the critical point and one expects novel modes and excitation channels to appear, which may signal possible unconventional critical behaviour.

To describe the low-energy excitation spectra, we apply the slave boson formalism described in Ref.~\cite{Frerot_Roscilde} to the cluster framework used in this work.
This and similar approaches \cite{Stoof2003, huber2007, huerga2013, krutitsky2016ultracold, DiLibertoMenotti, Wang_topoHiggs2021} have proven to accurately capture the excitation gaps observed experimentally \cite{endres2012higgs} and can include quantum fluctuations in a controlled manner \cite{caleffi2020, Frerot_Roscilde}. 
Within this framework, for each plaquette we introduce slave boson operators $\hat{\beta}^{}_{p,{\mathbf{n}}}$, spanning the plaquette Hilbert space $| n_{1}^{} n_{2}^{} n_{3}^{} n_{4}^{} \rangle_p^{} \equiv | \mathbf{n} \rangle_p^{} = \hat{\beta}_{p,\mathbf{n}}^{\dagger} |0_{SB}^{}\rangle$, where $|0_{SB}^{}\rangle$ is a fictitious slave boson vacuum and $\sum_{\mathbf n} \hat{\beta}_{p,\mathbf{n}}^{\dagger}\hat{\beta}_{p,\mathbf{n}}^{}=1$.
The diagonalization of the cluster mean-field Hamiltonian $\hat{\mathcal H}_\text{MF}^{(p)}$ provides a set of plaquette normal modes described by the unitary transformation  $U_{m,{\mathbf{n}}}^{(p)}$ that we use to rotate the slave boson operators via the linear transformation 
$\hat{\gamma}^{\dagger}_{p,m} = \sum_{{\mathbf{n}}} U_{m,{\mathbf{n}}}^{(p)} \hat{\beta}^{\dagger}_{p,{\mathbf{n}}}$,
where $m$ labels each plaquette normal mode with increasing energy. 
The bosonic creation operator can then be expressed as
\begin{equation}
\hat{b}_{i,p}^{\dagger} = \sum_{{\mathbf{n}}{\mathbf{n}'}} \langle \mathbf{n} | \hat{b}_{i,p}^{\dagger} | \mathbf{n'} \rangle\hat{\beta}^{\dagger}_{p,{\mathbf{n}}} \hat{\beta}^{}_{p,{\mathbf{n}'}} = \sum_{n ,m} \hat{\gamma}^{\dagger}_{p,n}\Tilde{F}^{i,p}_{nm}\hat{\gamma}^{}_{p,m}\,.
\end{equation}
Within the slave boson approach, we approximate $\hat{\mathcal H}_\text{MF}^{(p)}$ by condensing $\hat{\gamma}_{p,m=0}^{(\dagger)} \approx 1$ and by treating $\hat{\gamma}_{p,m \neq 0}^{(\dagger)}$ as weak quantum fluctuations up to quadratic order. 
In momentum space, we obtain\cite{SuppMat} the quadratic Hamiltonian 
\begin{equation}
\hat{\mathcal{H}}\approx\mathcal{\hat{H}}^{(2)}_{} = \frac{1}{2} \sum_{\textbf{k}} ( \boldsymbol{\hat{\gamma}}^{\dagger}_{\textbf{k}}, \boldsymbol{\hat{\gamma}}_{\textbf{-k}}^{})
H_{\mathbf{k}}^{}
(\boldsymbol{\hat{\gamma}}_{\textbf{k}}^{}, \boldsymbol{\hat{\gamma}}^{\dagger}_{\textbf{-k}})^{T}\,, 
\end{equation}
where $H_{\mathbf{k}}$ is a $(2d-2) \times (2d-2)$ BdG Hamiltonian matrix,  $\boldsymbol{\hat{\gamma}}_{p}= 1/\sqrt{N} \sum_{\textbf{k}} e^{i\textbf{k}\cdot\textbf{r}_{p}}\boldsymbol{\hat{\gamma}^{}_{\textbf{k}}}$ and $\boldsymbol{\hat{\gamma}}_{p} = ( \hat{\gamma}_{p,1}^{}, \dots, \hat{\gamma}_{p,n}^{} \dots \hat{\gamma}_{p,d}^{})^{T}$.
As shown in Fig.~\ref{Fig4}(a)-(c) near the critical point of the transition at $\nu=1$, the excitation spectra obtained from $H_{\mathbf{k}}$ display several low-energy branches in the superfluid phase. 
The gapless branch corresponds to the Goldstone (phase) mode.
However, other low-energy gapped modes appear.
The analysis of the phase and amplitude character of the order parameter, obtained via the flatness $\mathcal{F}_{\alpha,\mathbf{k}}$~\cite{endres2012higgs, DiLibertoMenotti,SuppMat}, reveals that two of the gapped excitations (H1 and H2) are amplitude (Higgs) modes with $\mathcal{F}_{\alpha,\mathbf{k}}\approx 1$ corresponding to oscillations of the superfluid density and only one softening at the transition, see Fig.~\ref{Fig4}(d), whereas only the lowest mode is a phase (Goldstone) mode ($\mathcal{F}_{\alpha,\mathbf{k}}\approx -1$). 
This result is in contrast to the more conventional behavior of Mott-superfluid transitions with only a single Higgs excitation. 
Two other dispersive modes appear with neither amplitude nor phase character. 
As shown in Fig.~\ref{Fig4}(b), these modes are particle or hole excitations as indicated by computing the density variation carried by each of them and their particle-hole character, see Ref.~\cite{DiLibertoMenotti}.
The chiral nature of the excitations can be revealed by estimating the loop current for each branch $\langle\hat{\mathcal L}\rangle_\alpha$ as compared to the ground state value $\langle\hat{\mathcal L}\rangle$, as shown in Fig.~\ref{Fig4}(c).
The two Higgs modes both carry a finite loop current. 
Sufficiently within the superfluid phase its value for $\mathbf k=0$ tends to the effective theory expectation $\langle \hat{\mathcal{L}}\rangle_{\rm{H1,H2}}  \rightarrow \pm\sqrt{2}$ (per particle), see Fig.~\ref{Fig4}(d). 
Crucially, the lowest-energy Higgs mode (H1) exhibits a loop current of opposite sign as compared to the ground state one, thus bringing a finite \emph{large} differential value $|\Delta \mathcal L_\rm{H1} | \equiv |\langle\hat{\mathcal L}\rangle_{\rm H1} -\langle\hat{\mathcal L}\rangle|\approx 2\sqrt 2$ per particle, and manifesting a chiral nature. 
This hierarchy is motivated by the fact that the nearby insulating phase does not break TRS, and the lowest Higgs H1 mode is there to carry away the ground-state loop current across the transition.
Notice that this multi-Higgs structure only appears near the $\nu=1$ critical point, whereas at fractional fillings a more conventional picture with a single non-chiral Higgs mode ($\Delta \mathcal L_\rm{H}\approx 0$) occurs \cite{SuppMat}. 

In order to probe the lowest Higgs chiral mode H1, we perform a quench by adding a weak staggered flux perturbation $\pm\delta\theta$ \cite{aidelsburger2011}, which preserves U(1) and explicitly breaks TRS. 
As shown in Fig.~\ref{Fig4}(e), this selectively excites coherent in-phase oscillations of loop current and order parameter. 
A Fourier spectral analysis, Fig.~\ref{Fig4}(f), reveals a dominant peak corresponding to the chiral Higgs mode H1, negligible excitation of H2 and no particle (hole) weight, consistent with the fact that the quench preserves the global U(1) symmetry \cite{podolsky2011, endres2012higgs}. 
Similar results are also obtained by the more traditional quench of the hopping amplitude $J$ \cite{endres2012higgs}.
The visibility of the Higgs mode via current oscillations makes its detection relevant in light of the recently developed protocol for (local/site-resolved) current measurement in optical lattices \cite{impertro2024}.

\section{Discussion and conclusions}
\label{sec:V}
In this work we have shown that interacting bosons on a dimerized flux-frustrated lattice display distinct quantum phases that spontaneously break TRS manifesting in vortex Mott and superfluid phases.
The transition at integer filling breaks TRS and U(1) symmetries, and the excitation spectrum in the chiral superfluid phase consists of two Higgs quasiparticles, including a chiral one. 
Our results provide an example of a strongly correlated system with nontrivial quasiparticles that is especially relevant in the context of very recent experimental developments employing dimerized lattices with synthetic gauge fields \cite{impertro2025} and the corresponding direct measurement of the current patterns.
Our work opens several questions concerning, for example, the decay and lifetime of the quasiparticles here identified \cite{podolsky2011}, the protocol to selectively excite the second Higgs mode, the effective quantum field-theory description of the multi-Higgs structure in the presence of coexisting orders,  the critical properties of the $\nu=1$ transition \cite{sachdev2023quantum} to be investigated with other methods, as for example large-scale Quantum Monte Carlo techniques \cite{Pollet2012}. 

Further work will clarify the possibility that the two symmetries ($\mathbb{Z}_2^T$ and U(1)) are broken separately at integer filling through an intermediate phase in a narrow parameter region, as identified via refined numerical DMRG analyses in related non-dimerized models \cite{Dhar2012, greschner2013, altman2014}.
Further investigations are needed to elucidate the interplay between topology and interactions in the BBH model at different fillings \cite{Bibo2020, Cuadra2022}, as for the SSH case \cite{Grusdt_topo}, including the possibility to stabilize topological phases via interactions \cite{Yao2024}.
In addition, it would be interesting to explore effects beyond  strong dimerization that are not captured by our analysis and the role played by gauge symmetries \cite{Burgher2025} and chiral bound pairs \cite{Stepanenko2024} in transport phenomena.
Our work paves the way to the identification of other unconventional critical points in chiral systems \cite{altman2014}, including those connected to deconfined critical points \cite{Polini2005, senthil2004deconfined, senthil2024deconfined}.

\section{Acknowledgements}
The authors thank N. Goldman, M. Morgavi, S. Montangero and L. Pave\v{s}i\'{c} for fruitful discussions.
The exact diagonalisation calculations were performed with \texttt{QuSpin}~\cite{Quspin}.
This work has received funding under the Horizon Europe programme HORIZON-CL4-2022-QUANTUM-02-SGA via the project 101113690 (PASQuanS2.1) and has been supported by the INFN project Iniziativa Specifica IS-Quantum, the Italian Ministry of University and Research via the Excellence grant 2023-2027 ``Quantum Frontiers",  the Rita Levi-Montalcini program.
The authors also acknowledge computational resources by Cloud Veneto and INFN Padova HPC cluster, funded by the NextGenerationEU Terabit project on PNRR – Avviso n. 3264 ``per il Rafforzamento e creazione di Infrastrutture di Ricerca”, Missione 4, ``Istruzione e
Ricerca”.

\bibliography{bibliography}

\clearpage
\onecolumngrid

\clearpage

\newcommand{\beginsupplement}{%
        \setcounter{equation}{0}
        \renewcommand{\theequation}{S\arabic{equation}}%
        \setcounter{figure}{0}
        \renewcommand{\thefigure}{S\arabic{figure}}%
     }
     
\beginsupplement

\begin{center}
{\large\bfseries Supplemental Material for}\\[0.5em]
{\Large\bfseries Emergent chiral Higgs mode in $\pi$-flux frustrated lattices}\\[1em]

M. Lanaro$^{1,2}$, L. Maffi$^{1,2}$, and M. Di Liberto$^{1,2,3}$\\[0.5em]

{\small $^{1}$Dipartimento di Fisica e Astronomia ``G. Galilei", Università degli Studi di Padova, I-35131 Padova, Italy\\ $^{2}$INFN Istituto Nazionale di Fisica Nucleare, Sezione di Padova, I-35131 Padova, Italy\\ $^{3}$Padua Quantum Technologies Research Center, via Marzolo 8, I-35131 Padova, Italy} \end{center}

\vspace{1em}

\section*{Cluster mean-field approximation.}
The Hamiltonian of our 2D system can be written as $\hat{\mathcal{H}} = \hat{\mathcal{H}}_{J}^{} + \hat{\mathcal{H}}_{\text{onsite}}^{} +\hat{\mathcal{H}}_{J'}^{} $, where
\begin{equation}
\begin{split}
    \hat{\mathcal{H}}_{J}^{} &= -J \sum_{p}\left( e^{i\pi} \hat{b}^{\dagger}_{1,p} \hat{b}_{2,p}^{} + \hat{b}^{\dagger}_{2,p} \hat{b}_{4,p}^{} + \hat{b}^{\dagger}_{4,p} \hat{b}_{3,p}^{} + \hat{b}^{\dagger}_{3,p} \hat{b}_{1,p}^{} + \text{h.c.} \right) \,, \\
    \hat{\mathcal{H}}_{J'}^{} &= -J'\sum_{p}\left( e^{i\pi} \hat{b}^{\dagger}_{2,p} \hat{b}_{1,p+\hat{\textbf{u}}_{y}}^{} + \hat{b}^{\dagger}_{4,p} \hat{b}_{3,p+\hat{\textbf{u}}_{y}}^{} +  \hat{b}^{\dagger}_{4,p} \hat{b}_{2,p+\hat{\textbf{u}}_{x}}^{} +  \hat{b}^{\dagger}_{3,p} \hat{b}_{1,p+\hat{\textbf{u}}_{x}}^{} + \text{h.c.} \right) \,, \\
    \mathcal{\hat{H}}_{\text{onsite}} &= \frac{U}{2} \sum_{p}\sum_{i=1}^{4} \hat{n}_{i,p}^{} (\hat{n}_{i,p}^{} -1 ) - \mu \sum_{p}\sum_{i=1}^{4} \hat{n}^{}_{i,p} \,,
\end{split}
\end{equation} 
where $(\hat{\mathbf{u}}_{x}, \hat{\mathbf{u}}_{y})$ are unit vectors along the $x$ and $y$ directions.
We solve the Hamiltonian $\hat{\mathcal H}$ within a cluster Gutzwiller ansatz approximation, namely a mean-field ansatz for plaquettes,
\be 
|\Phi \rangle_{\text{cl}} \equiv \bigotimes_{p} \left( \sum_{n_{1}^{}n_{2}^{}n_{3}^{}n_{4}^{}} c^{(p)}_{n_{1}^{}n_{2}^{}n_{3}^{}n_{4}^{}}|n_{1}^{}n_{2}^{}n_{3}^{}n_{4}^{}\rangle_{p}^{} \right) \,,
\ee
from which we can rewrite $\hat{\mathcal{H}}$ as
\be 
\mathcal{\hat{H}} \approx \mathcal{\hat{H}}^{}_{\text{cl}} + \mathcal{\hat{H}}' \ ,
\ee
with $\mathcal{\hat{H}}^{}_{\text{cl}}=\mathcal{\hat{H}}^{}_{J}+\mathcal{\hat{H}}^{}_{\text{onsite}}$ and the coupling between different plaquettes taking the form 
\be
\hat{\mathcal{H}}_{J'}^{} \rightarrow \hat{\mathcal{H}}' = -\sum_{\substack{\langle p,q \rangle \\ \langle i,j \rangle }} \left[J'_{ij} \left(\hat{b}_{i,p}^\dagger \psi^{j,q}_{} -(\psi^{j,q})^*\psi^{i,p} \right) +\text{h.c.} \right] \,,
\ee
where $\psi^{i,p}\equiv \langle \hat b_{i,p}\rangle$ is the order parameter for the $i$-th site of the $p$-th plaquette. 
This allows us to write $\hat{\mathcal H}\approx\sum_p \hat{\mathcal H}_\text{MF}^{(p)}$, where $\hat{\mathcal H}_\text{MF}^{(p)}$ contains only operators acting on a single plaquette $p$. 
The local basis here takes the form 
$| \mathbf{n} \rangle = | n_{1}^{} n_{2}^{} n_{3}^{} n_{4}^{} \rangle$, where the different $(n_{1}^{}, n_{2}^{}, n_{3}^{}, n_{4}^{}) \equiv \mathbf{n}$ represent the occupation number on each site of the sublattice of the plaquette. We truncate to $n_i < n_{\text{max}}$, with $n_{\text{max}}=5$ and a corresponding local Hilbert space dimension $d=n_{\text{max}}^4$.
The Fock states are generated by a fictitious slave boson vacuum, $|\mathbf{n}\rangle_{p} = \hat{\beta}_{p,\mathbf{n}}^{\dagger} |0_{SB}\rangle_{p}^{}$, with $\sum_{\mathbf n} \hat{\beta}_{p,\mathbf{n}}^{\dagger}\hat{\beta}_{p,\mathbf{n}}^{}=1$.
We rotate the $\hat{\beta}$ operators to new $\hat{\gamma}$ operators via
\begin{equation}
    \hat{\gamma}^{\dagger}_{p,
    n} = \sum_{ \mathbf{n} } U_{n,\mathbf{n} }^{(p)} \hat{\beta}_{p,
    \mathbf{n} }^{\dagger} \ ,
\end{equation}
with $U_{n, \mathbf{n}}^{(p)}$ the unitary transformation that diagonalizes the mean-field solution of $\hat{\mathcal H}$, namely $\hat{\mathcal H}_\text{MF}^{(p)}$, and whose columns are the mean-field eigenvectors $\mathbf{\Psi}^{(p)}=(\psi^{p}_{0}, \dots, \psi^{p}_{n}, \dots, \psi^{p}_{d})$ 
\begin{equation}
    \langle \mathbf{n} | \Psi^{(p)}_{n} \rangle = U_{n,
    \mathbf{n}}^{(p)} \, .
\end{equation}
In this way, we get a Hamiltonian with quartic terms in $\hat{\gamma}$ for the hopping part $\hat{\mathcal{H}}_{J'}^{}$, and quadratic terms (within the plaquette) for the rest.
In terms of the rotated $\hat{\gamma}$ operators, the Hamiltonian takes the form 
\begin{equation}
    \hat{\mathcal{H}} = -J' \sum_{\langle p,q \rangle} \sum_{\langle i,j \rangle} \left( \boldsymbol{\hat{\gamma}}_{p}^{\dagger} \Tilde{F}^{i,p}\boldsymbol{\hat{\gamma}}_{p} \boldsymbol{\hat{\gamma}}_{q}^{\dagger} [\Tilde{F}^{j,q}]^{\dagger} \boldsymbol{\hat{\gamma}}_{q} + \text{h.c}   \right) + \sum_{p} \boldsymbol{\hat{\gamma}}_{p}^{\dagger} \Tilde{G} \boldsymbol{\hat{\gamma}}_{p} \label{SB_Hamiltonian} \ ,
\end{equation}
where 
\begin{align}
    \hat{\gamma}_{p,n}^{\dagger} &= \sum_{ \mathbf{n}} \langle \mathbf{n} | \Psi_{n}^{(p)} \rangle \hat{\beta}_{p,\mathbf{n}}^{\dagger} \,, \\
    \boldsymbol{\hat{\gamma}}_{p} &= (\hat{\gamma}_{p,0}, ..., \hat{\gamma}_{p,n},...,\hat{\gamma}_{p,d})^{T} \,, \\
    \Tilde{F}^{i,p}_{nm} &= \langle \Psi^{(p)}_{n}|\hat{b}^{\dagger}_{i,p}| \Psi_{m}^{(p)} \rangle \,, \\
    \Tilde{G}_{n m} &= \langle \Psi_{n}^{(p)}| \hat{\mathcal{H}}_{J}^{} + \hat{\mathcal{H}}_{\text{onsite}}^{} | \Psi_{m}^{(p)} \rangle \, .
\end{align}
Notice that in the definition of $\tilde G_{nm}$ we have included also hopping processes within the plaquette, which marks a difference with the treatment discussed in Ref.~\cite{Frerot_Roscilde} and allows us to introduce intra-plaquette correlations. 
In order to proceed further with the calculations we perform now a quadratic approximation around the mean field minimum, condensing the $\hat{\gamma}_{p,0}^{}$ bosons ($\hat{\gamma}_{p,0}^{\dagger} \approx 1$) and treating the other bosonic fields as perturbations, i.e. $\hat{\gamma}^{\dagger}_{p,n \neq 0 } \hat{\gamma}^{}_{p,n \neq 0 }  \ll 1$. In this way (Bogoliubov shift)
\begin{equation} \label{Bogoliubov_shift}
    \hat{\gamma}^{(\dagger)}_{p,0} \approx \sqrt{1 - \sum_{n \neq 0}   \hat{\gamma}_{p,n}^{\dagger} \hat{\gamma}_{p,n}^{} } \approx 1 - \frac{1}{2} \sum_{n \neq 0} \hat{\gamma}_{p,n}^{\dagger} \hat{\gamma}_{p,n}^{} + O(\hat{\gamma}^{4}) \ .
\end{equation}
\begin{figure}
    \centering
    \includegraphics[width=0.98\linewidth]{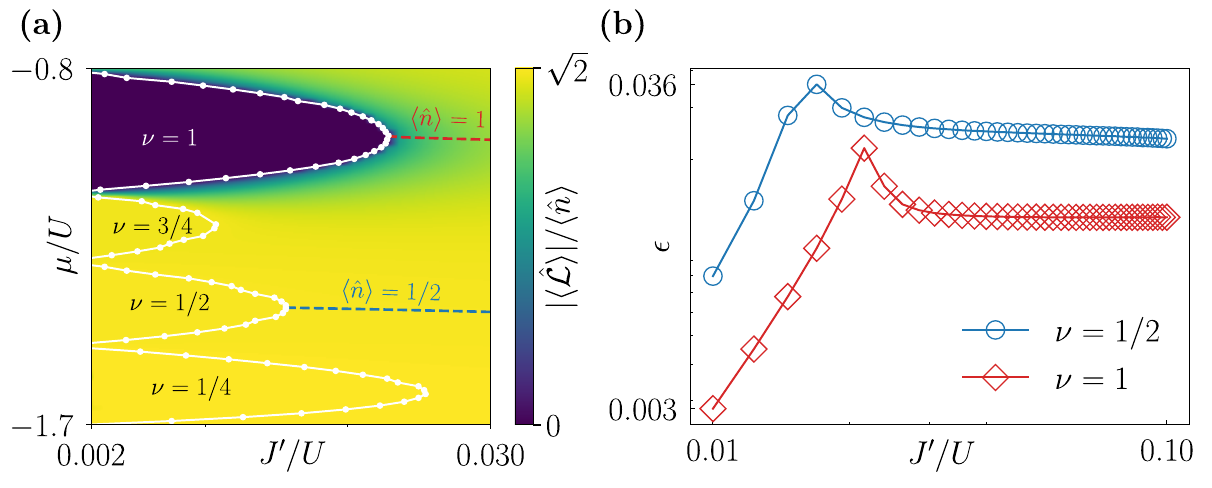}
    \caption{\textbf{(a)} Phase diagram along the cut at fixed $J/U=1.2$. White dots mark the transition points from the V-SF (right) to the (V-)MI (left), while the dotted curves show the constant filling lines for $\nu=1$ (red) and $\nu=1/2$ (blue). The vertical axis shows the ground-state loop current. The lobe at integer filling exhibits zero loop current, while the fractional-filling lobes correspond to V-MI phases with finite current. \textbf{(b)} Control parameter at integer (red) and half (blue) filling for $J/U=1.2$ as a function of $J ' /U$. The control parameter displays a peak at the transition point and remains below $\epsilon \sim 0.04$.}
    \label{Control_param_phase_diag}
\end{figure}
In order for this to be consistent, we must verify that the average values of the non-condensed slave bosons remain small in the renormalized ground-state, i.e.
\begin{equation}
    \epsilon = \sum_{n>0} \langle \hat{\gamma}_{p,n}^{\dagger}    \hat{\gamma}_{p,n}^{} \rangle \ll 1 \ ,
\end{equation}
as we can see in Fig. \ref{Control_param_phase_diag}(b) for the cut with fixed $J/U=1.2$ (phase diagram shown in Fig. \ref{Control_param_phase_diag}(a)). In this figure, the peak in the control parameter appears at the critical point for filling $\nu=1$ and $\nu=1/2$, but we can observe how $\epsilon$ remains always small. In particular, when $J'/J=0.03$ and $\nu=1$ we find $\epsilon \sim 0.033$ at the transition point.
With this, the field operator $\hat{b}_{i,p}^\dagger$ is expressed as
\begin{equation}\begin{split}
 \hat{b}_{i,p}^{\dagger} &= \sum_{n m} \hat{\gamma}_{p,n}^{\dagger} \Tilde{F}_{nm}^{i,p}\hat{\gamma}_{p,m} \\
 &= \hat{\gamma}_{p,0}^{\dagger} \Tilde{F}^{i,p}_{00} \hat{\gamma}_{p,0}
 + \sum_{n>0}\big(\hat{\gamma}_{p,0}^{\dagger} \Tilde{F}^{i,p}_{0n} \hat{\gamma}_{p,n}
 + \hat{\gamma}_{p,n}^{\dagger} \Tilde{F}^{i,p}_{n0} \hat{\gamma}_{p,0}\big) + \sum_{n,m>0} \hat{\gamma}_{p,n}^{\dagger} \Tilde{F}^{i,p}_{nm} \hat{\gamma}_{p,m} \\
 &\approx \Tilde{F}_{00}^{i,p} \Big( 1- \sum_{n>0} \hat{\gamma}_{p,n}^{\dagger} \hat{\gamma}_{p, n} \Big)
 + \sum_{n>0}(\Tilde{F}^{i,p}_{0 n}\hat{\gamma}_{p, n}
 + \hat{\gamma}_{p, n}^{\dagger} \Tilde{F}^{i,p}_{n 0}) + \sum_{n, m > 0} \hat{\gamma}^{\dagger}_{p, n} \Tilde{F}^{i,p}_{n m}\hat{\gamma}_{p,m} + O(\hat{\gamma}^{3}) \ .
\end{split}
\label{btogamma}
\end{equation}
and similarly for the $\hat{b}_{i,p}$ operator. 
Notice that 
\begin{equation}
\psi_{i,p} = \mv{\hat b_{i,p}} = [\tilde F_{00}^{i,p}]^*
\end{equation}
It is possible to verify that
\begin{equation}
\begin{split}
    [\Tilde{F}^{i,p}]^*_{m n} &= [\Tilde{F}_{n m}^{i,p}]^{\dagger} \ , \\
    [\Tilde{F}_{m n}^{i,p}]^* = \left[ \langle \Psi_{m}^{(p)}| \hat{b}_{i,p}^{\dagger} | \Psi_{n}^{(p)} \rangle \right]^{*} &= \left[ \langle \hat{b}_{i,p}\Psi_{m}^{(p)}|  \Psi_{n}^{(p)} \rangle\right]^{*} = \langle\Psi_{n}^{(p)}|\hat{b}_{i,p}|\Psi^{(p)}_{m} \rangle \ .
\end{split}
\end{equation}
In order to treat bilinears, for example $\hat{b}_{i,p}^{\dagger}\hat{b}_{ j,q}^{}$ for different plaquettes, we expand in terms of $\Tilde{F}$ and $\hat{\gamma}$
\be\label{bibj_SB_different_plaq}
\begin{split}
\hat{b}_{i,p}^{\dagger}\hat{b}_{ j,q} = &\left[\sum_{n>0} \Tilde{F}_{00}^{i,p} \left( 1- \hat{\gamma}_{p, n}^{\dagger} \hat{\gamma}_{p, n} \right) + \sum_{n>0}  (\Tilde{F}_{0 n}^{i,p} \hat{\gamma}_{p, n} + \hat{\gamma}_{p, n}^{\dagger} \Tilde{F}_{n 0}^{i,p}) \sum_{n, m>0} \hat{\gamma}^{\dagger}_{p, n} \Tilde{F}^{i,p}_{n m}\hat{\gamma}_{p, m}\right] \\ \times &\left[\sum_{n'>0} [\Tilde{F}_{00}^{j,q}]^* \left( 1- \hat{\gamma}_{q, n'}^{\dagger} \hat{\gamma}_{q, n'} \right) + \sum_{n'>0} ([\Tilde{F}_{0 n'}^{j,q}]^* \hat{\gamma}_{q, n'}^{\dagger} + [\Tilde{F}_{n' 0}^{j,q}]^* \hat{\gamma}_{q, n'} ) \sum_{n', m'>0}  \hat{\gamma}^{\dagger}_{q, m'} [\Tilde{F}_{n' m'}^{j,q}]^* \hat{\gamma}_{q, n'} \right] = 
\\ =& \Tilde{F}_{00}^{i,p}[\Tilde{F}_{00}^{j,q}]^* + \\-& \Tilde{F}_{00}^{i,p}[\Tilde{F}_{00}^{j,q}]^*(\sum_{n>0} \hat{\gamma}_{p, n}^{\dagger} \hat{\gamma}_{p, n} + \sum_{n'>0} \hat{\gamma}_{q, n'}^{\dagger} \hat{\gamma}_{q, n'}) + [\Tilde{F}_{00}^{j,q}]^*\sum_{n,m>0} \hat{\gamma}_{p, n}^{\dagger} \Tilde{F}^{i,p}_{n m} \hat{\gamma}_{p, m} + \Tilde{F}^{i,p}_{00} \sum_{n',m'>0} \hat{\gamma}_{q m'}^{\dagger} [\Tilde{F}_{n' m'}^{j,q}]^* \hat{\gamma}_{q, n'} +\\+& \Tilde{F}_{00}^{i,p}\sum_{n'>0} ( [\Tilde{F}_{0 n'}^{j,q}]^* \hat{\gamma}_{q, n'}^{\dagger} + [\Tilde{F}_{n' 0}^{j,q}]^* \hat{\gamma}_{q, n'} ) +[\Tilde{F}_{00}^{j,q}]^* \sum_{n>0} ( \Tilde{F}^{i,p}_{0 n} \hat{\gamma}_{p, n} + \hat{\gamma}_{p, n}^{\dagger} \Tilde{F}^{i,p}_{n 0}) + \\
+& \sum_{n, n'>0} (   \Tilde{F}^{i,p}_{0 n} \hat{\gamma}_{p, n} \hat{\gamma}_{q, n'}^{\dagger} [\Tilde{F}_{0 n'}^{j,q}]^* + \hat{\gamma}_{p, n}^{\dagger} \Tilde{F}^{i,p}_{n 0} [\Tilde{F}_{n' 0}^{j,q}]^* \hat{\gamma}_{q, n'})+ \\ +& \sum_{n, n'>0} (\hat{\gamma}_{p, n}^{\dagger} \Tilde{F}^{i,p}_{n 0} \hat{\gamma}_{q, n'}^{\dagger} [\Tilde{F}_{0 n'}^{j,q}]^* + \Tilde{F}^{i,p}_{0 n} \hat{\gamma}_{p, n} [\Tilde{F}_{n' 0}^{j,q}]^* \hat{\gamma}_{q, n'}) \,.
\end{split}
\ee 
With this expression we can now work on Eq.~\eqref{SB_Hamiltonian} and reduce it to
\begin{equation}
\begin{split}
\mathcal{\hat{H}}^{(0)} &= -\sum_{\mv{p,q}}\sum_{\mv{i,j}}J'_{ij}\tilde F_{00}^{i,p}[\tilde F_{00}^{j,q}]^* + N\tilde G_{00} \ ,\\
\mathcal{\hat{H}}^{(1)} &= \sum_{p}\sum_{n>0} \hat{\gamma}_{p, n} \langle \Psi_{0} | \mathcal{\hat{H}}_{MF}^{(p)} | \Psi_{n} \rangle + \text{h.c}
\ ,\\
\mathcal{\hat{H}}_{\text{plaq}} &= \sum_{p} \sum_{n, m >0} \hat{\gamma}_{p, n}^{\dagger} A^{(0)}_{n m} \hat{\gamma}_{p, m}
\ ,\\
\mathcal{\hat{H}}_{\text{hop}} &= \sum_{\langle p,q \rangle} \sum_{n , m >0} \hat{\gamma}_{p, n}^{\dagger} A_{n m}^{(1)(p,q)}
\hat{\gamma}_{q,m}\ , \\
\mathcal{\hat{H}}_{\text{pairs}} &= \frac{1}{2} \sum_{\langle p,q \rangle} \sum_{n,m>0} \hat{\gamma}_{p, n} B^{(p,q)}_{n m} \hat{\gamma}_{q,m} + \text{h.c}\,,
\end{split}
\end{equation}
where $N$ is the number of sites.
The matrices $A$ and $B$ take the form
\begin{equation}
\begin{split}
    A_{n m}^{(0)} &= -\delta_{nm} \langle \Psi_0^{(p)} | \mathcal{H}_{MF}^{(p)} | \Psi_{0}^{(p)} \rangle + \langle \Psi_{n}^{(p)} | \mathcal{H}_{MF}^{(p)} | \Psi_{m}^{(p)} \rangle = \delta_{n m} (\epsilon_{n} - \epsilon_{0}) \ ; \\
    A_{n m}^{(1)(p,q)}  &= - \sum_{\langle i,j \rangle}J'_{ij}(\Tilde{F}_{n 0}^{i,p} [\Tilde{F}_{m 0 }^{j,q}]^* + \Tilde{F}_{0 m}^{i,p} [\Tilde{F}_{0 n}^{j,q}]^*) \ ; \\
    B_{n m}^{(p,q)} &= -\sum_{\langle i,j \rangle}J'_{ij}( \Tilde{F}_{0  n}^{i,p} [\Tilde{F}_{m 0}^{j,q}]^* + \Tilde{F}_{ 0 m }^{i,p} [\Tilde{F}_{n 0}^{j,q}]^*) \ .
\end{split}
\end{equation}
Since $\hat{\mathcal{H}}_{0}$ is an offset, while $\hat{\mathcal{H}}_{1}$ vanishes, the non-vanishing part of the Hamiltonian is 
\begin{equation}
    \mathcal{\hat{H}}^{(2)} = \mathcal{\hat{H}}_{\text{plaq}} + \mathcal{\hat{H}}_{\text{hop}} + \mathcal{{\hat{H}}}_{\text{pairs}} \ .
\end{equation}
We focus on our specific case of a cluster of 4 sites in a 2D plaquette configuration, for which we want to find the Fourier transform in order to get the excitation spectrum. We distinguish between vertical hoppings ($U-D$, for example between site (2,p) and (1,p+1)) and horizontal hoppings ($R-L$, for example between site (4,p) and (2,p+1)). The $U-D$ part of $A^{(1)}$ is
\begin{equation}\label{A1_equation_SB}
    A^{(1)(p,p+1)_{UD}}_{n m} =  \hat{\gamma}_{p,n}^{\dagger} \underbrace{\left( \Tilde{F}_{n 0}^{U,p}[\Tilde{F}_{m0}^{D,p+1}]^* + [\Tilde{F}_{0n}^{U,p}]^*\Tilde{F}_{0m}^{D,p+1}\right)}_{A^{(DU)}}\hat{\gamma}_{p+1,m} 
    + \hat{\gamma}_{p+1,n}^{\dagger} \underbrace{ \left( [\Tilde{F}_{0n }^{D,p+1}]^*\Tilde{F}_{0m}^{U,p} + \Tilde{F}_{n0}^{D,p+1}[\Tilde{F}_{m0}^{U,p}]^*\right)}_{A^{(UD)}}\hat{\gamma}_{p,m} \, .
\end{equation}
With the same procedure we have, for the left-right interaction
\begin{equation}
     A^{(1)(p,p+1)_{LR}}_{n m} =  \hat{\gamma}_{p,n}^{\dagger} \underbrace{\left( \Tilde{F}_{n 0}^{R,p}[\Tilde{F}_{m0}^{L,p+1}]^* + [\Tilde{F}_{0n}^{R,p}]^* \Tilde{F}_{0m}^{L,p+1}\right)}_{A^{(LR)}}\hat{\gamma}_{p+1, m} 
    + \hat{\gamma}_{p+1,n}^{\dagger} \underbrace{\left( [\Tilde{F}_{0n }^{L,p+1}]^*\Tilde{F}_{0m}^{R,p} + \Tilde{F}_{n0}^{L,p+1}[\Tilde{F}_{m0}^{R,p}]^*\right)}_{A^{(RL)}}\hat{\gamma}_{p,m} \ .
\end{equation}
We can notice that 
\begin{equation}
    \left(A\right)_{n m} = \left(A\right)^{\dagger}_{n m} \ ,
\end{equation}
both for the left-right and up-down interaction, so that we define
\begin{align}
    \textrm{A}_{1} &= \textrm{A}^{DU} + \textrm{A}^{LR} \\
    \textrm{A}_{2} &= \textrm{A}^{UD} + \textrm{A}^{RL} \ .
\end{align}
The Fourier transform of $\boldsymbol{\gamma}_{p}$ is
\begin{equation}\label{FT_gamma_SB}
    \boldsymbol{\hat{\gamma}}_{p}= \frac{1}{\sqrt{N}} \sum_{\textbf{k}} e^{i\mathbf{k \cdot r}_{p}} \boldsymbol{\hat{\gamma}_{\textbf{k}}} \ ,
\end{equation}
with $\textbf{r}_{q} = \textbf{r}_{p} + \boldsymbol{\delta}$. $\boldsymbol{\delta}=(\mathbf{\hat{u}}_{x}, \mathbf{\hat{u}}_{y})a$ and $a$ is the lattice spacing. If we insert this in \eqref{A1_equation_SB}
\begin{align*}
    A^{(1)}_{\textbf{k}} &= \frac{1}{N}\sum_{p} \sum_{\textbf{k}\textbf{k}'} \left[ e^{-i\mathbf{k \cdot r}_{p}} \boldsymbol{\hat{\gamma}}_{\textbf{k}}^{\dagger} \textrm{A}_{1} e^{i\textbf{k}'\cdot (\textbf{r}_{p} + \boldsymbol{\delta})} \boldsymbol{\hat{\gamma}}_{\textbf{k}'} + e^{-i\textbf{k}\cdot (\textbf{r}_{p}+\boldsymbol{\delta})} \boldsymbol{\hat{\gamma}}_{\textbf{k}}^{\dagger} \textrm{A}_{2} e^{i\mathbf{k' \cdot r}_{p}} \boldsymbol{\hat{\gamma}}_{\textbf{k}'} \right] = \\
    & = \sum_{\textbf{k}} \left[e^{i\textbf{k}\cdot \boldsymbol{\delta}}\boldsymbol{\hat{\gamma}}_{\textbf{k}}^{\dagger} \textrm{A}_{1} \boldsymbol{\hat{\gamma}}_{\textbf{k}} + e^{-i\textbf{k}\cdot \boldsymbol{\delta}}\boldsymbol{\hat{\gamma}}_{\textbf{k}}^{\dagger} \textrm{A}_{2} \boldsymbol{\hat{\gamma}}_{\textbf{k}} \right] \ ,
\end{align*}
and symmetrizing
\begin{align*}
        A^{(1)}_{\textbf{k}} &= \frac{1}{2}\sum_{\textbf{k}} \left[e^{i\textbf{k}\cdot \boldsymbol{\delta}}\boldsymbol{\hat{\gamma}}_{\textbf{k}}^{\dagger} \textrm{A}_{1} \boldsymbol{\hat{\gamma}}_{\textbf{k}} + e^{-i\textbf{k}\cdot \boldsymbol{\delta}}\boldsymbol{\hat{\gamma}}_{\textbf{-k}}^{\dagger} \textrm{A}_{1} \boldsymbol{\hat{\gamma}}_{\textbf{-k}} + e^{-i\textbf{k}\cdot \boldsymbol{\delta}}\boldsymbol{\hat{\gamma}}_{\textbf{k}}^{\dagger} \textrm{A}_{2} \boldsymbol{\hat{\gamma}}_{\textbf{k}} + e^{i\textbf{k}\cdot \boldsymbol{\delta}}\boldsymbol{\hat{\gamma}}_{\textbf{-k}}^{\dagger} \textrm{A}_{2} \boldsymbol{\hat{\gamma}}_{\textbf{-k}} \right] =
        \\ & = \frac{1}{2} \sum_{\textbf{k}} \left[ \boldsymbol{\hat{\gamma}}_{\textbf{k}}^{\dagger} \left( e^{i\textbf{k}\cdot \boldsymbol{\delta}} \textrm{A}_{1} + e^{-i\textbf{k}\cdot \boldsymbol{\delta}}A_{2} \right) \boldsymbol{\hat{\gamma}}_{\textbf{k}} + \underbrace{\boldsymbol{\hat{\gamma}}_{\textbf{-k}}^{\dagger} \left( e^{-i\textbf{k}\cdot \boldsymbol{\delta}} \textrm{A}_{1} + e^{i\textbf{k}\cdot \boldsymbol{\delta}} \textrm{A}_{2} \right) \boldsymbol{\hat{\gamma}}_{\textbf{-k}}}_{= \boldsymbol{\hat{\gamma}}_{\textbf{-k}} \left( e^{-i\textbf{k}\cdot \boldsymbol{\delta}} \textrm{A}_{1} + e^{i\textbf{k}\cdot \boldsymbol{\delta}} \textrm{A}_{2} \right)^{\textrm{T}} \boldsymbol{\hat{\gamma}}_{\textbf{-k}}^{\dagger}}  \right] \ .
\end{align*}
We then define $A_{\textbf{k}} = A^{(0)} + A^{(1)}_{\textbf{k}}$.\newline

\begin{figure}
    \centering
    \includegraphics[width=0.98\linewidth]{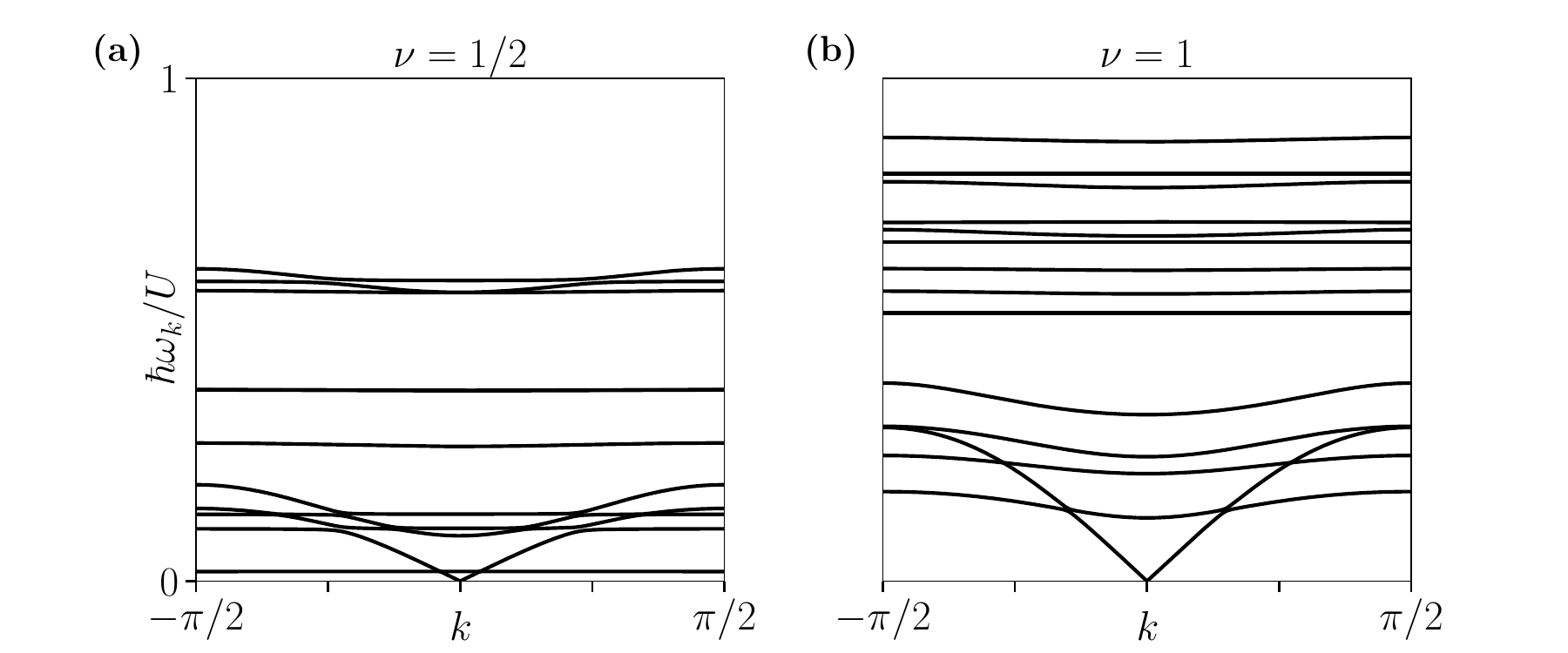}
    \caption{First excitations bands computed with the slave boson approach for \textbf{(a)} $\nu=1/2$ at $J'/U = 0.02$, $J/U = 0.67$, and $\mu/U = -0.67$, with $|\mathbf{k}_x| = |\mathbf{k}_y| = k$ and for \textbf{(b)} $\nu=1$ at $J'/U = 0.03$, $J/U = 1$, and $\mu/U = -0.68$, with $|\mathbf{k}_x| = |\mathbf{k}_y| = k$.} 
    \label{SM_fullSpectrum}
\end{figure}

Regarding the B term we have that 
\begin{equation}
    \textrm{B}^{(p,p+1)_{DU}}_{n m} = \hat{\gamma}_{p,n} \underbrace{\left( \Tilde{F}_{0n }^{U,p}[\Tilde{F}_{m0}^{D,p+1}]^* + [\Tilde{F}_{n0}^{U,p}]^*\Tilde{F}_{0,m}^{D,p+1}\right)}_{\textrm{B}^{DU}_{1}}\hat{\gamma}_{p+1,m} 
    + \hat{\gamma}_{p,n}^{\dagger} \underbrace{\left( \Tilde{F}_{n0 }^{U,p}[\Tilde{F}_{0m}^{D,p+1}]^* + [\Tilde{F}_{0n}^{U,p}]^*\Tilde{F}_{m0}^{D,p+1}\right)}_{\textrm{B}^{DU}_{2}}\hat{\gamma}_{p+1,m}^{\dagger} \ ,
\end{equation}
while for the left-right interaction we get
\begin{equation}
    \textrm{B}^{(p,p+1)_{LR}}_{n m} = \hat{\gamma}_{p,n} \underbrace{\left( \Tilde{F}_{0n }^{R,p}[\Tilde{F}_{m0}^{L,p+1}]^* + [\Tilde{F}_{n0}^{R,p}]^*\Tilde{F}_{0m}^{L,p+1}\right)}_{\textrm{B}^{LR}_{1}}\hat{\gamma}_{p+1,m} 
    + \hat{\gamma}_{p,n}^{\dagger} \underbrace{\left( \Tilde{F}_{n0 }^{R,p}[\Tilde{F}_{0m}^{L,p+1}]^* + [\Tilde{F}_{0n}^{R,p}]^*\Tilde{F}_{m0}^{L,p+1}\right)}_{\textrm{B}^{LR}_{2}}\hat{\gamma}_{p+1,m}^{\dagger} \ .
\end{equation}
We notice that
\begin{equation}
    \left(\textrm{B}^{}_1\right)_{n m} = \left(\textrm{B}^{}_2\right)_{n m}^{*} \ ,
\end{equation}
both for the left-right and up-down interaction, so that we define
\begin{equation}
\begin{split}
    \textrm{B}_{1}^{} &= \textrm{B}^{DU}_{1} + \textrm{B}^{LR}_{1} \\
    \textrm{B}_{2}^{} &= \textrm{B}^{DU}_{2} + \textrm{B}^{LR}_{2} \ .
\end{split}
\end{equation}
Taking the Fourier transform thanks to \eqref{FT_gamma_SB} leads to
\begin{equation}
\begin{split}
    \textrm{B}_{\textbf{k}}^{} &= \sum_{\textbf{k,k}'} \left[ 
    e^{i\mathbf{k \cdot r}_{p}} \boldsymbol{\hat{\gamma}}^{}_{\textbf{k}} \textrm{B}_{1}^{} e^{i\textbf{k}'(\textbf{r}_{p} + \boldsymbol{\delta})} \boldsymbol{\hat{\gamma}}_{\textbf{k}'}^{} + e^{-i\mathbf{k \cdot r}_{p}} \boldsymbol{\hat{\gamma}}_{\textbf{k}}^{\dagger} \textrm{B}_{2}^{} e^{i\textbf{k}'\cdot(\textbf{r}_{p} + \boldsymbol{\delta})} \boldsymbol{\hat{\gamma}}_{\textbf{k}'}^{\dagger}  \right] =  \\ &= \sum_{\textbf{k}} \left[ \boldsymbol{\hat{\gamma}}_{\textbf{k}}^{} \textrm{B}_{1}^{} \boldsymbol{\hat{\gamma}}^{}_{\textbf{-k}} e^{i\textbf{k}\cdot \boldsymbol{\delta}} + \boldsymbol{\hat{\gamma}}_{\textbf{k}}^{\dagger} \textrm{B}_{2}^{} \boldsymbol{\hat{\gamma}}_{\textbf{-k}}^{\dagger} e^{i\textbf{k}\cdot \boldsymbol{\delta}} \right] \ .
\end{split}
\end{equation}
With symmetrization we can get to
\begin{align}
    \textrm{B}_{\textbf{k}}^{} &= \frac{1}{2} \sum_{\textbf{k}} \left[ \boldsymbol{\hat{\gamma}}^{}_{\textbf{k}} \textrm{B}_{1}^{} \boldsymbol{\hat{\gamma}}^{}_{\textbf{-k}} e^{-i\textbf{k}\cdot \boldsymbol{\delta}} + \boldsymbol{\hat{\gamma}}^{}_{\textbf{-k}} \textrm{B}_{1}^{} \boldsymbol{\hat{\gamma}}^{}_{\textbf{k}} e^{i\textbf{k}\cdot \boldsymbol{\delta}} + \boldsymbol{\hat{\gamma}}_{\textbf{k}}^{\dagger} \textrm{B}_{2}^{} \boldsymbol{\hat{\gamma}}_{\textbf{-k}}^{\dagger} e^{i\textbf{k}\cdot \boldsymbol{\delta}}+ \boldsymbol{\hat{\gamma}}_{\textbf{-k}}^{\dagger} \textrm{B}_{2}^{} \boldsymbol{\hat{\gamma}}_{\textbf{k}}^{\dagger} e^{-i\textbf{k}\cdot \boldsymbol{\delta}} \right] \\ & 
    =\frac{1}{2} \sum_{\textbf{k}} \left[ \boldsymbol{\hat{\gamma}}_{\textbf{-k}}^{} \left(\textrm{B}_{1}^{\textrm{T}} e^{-i\textbf{k}\cdot \boldsymbol{\delta}} + \textrm{B}_{1}^{} e^{i\textbf{k}\cdot \boldsymbol{\delta}}\right)\boldsymbol{\hat{\gamma}}_{\textbf{k}}^{} + \boldsymbol{\hat{\gamma}}_{\textbf{k}}^{\dagger} \left(\textrm{B}_{2}^{} e^{i\textbf{k}\cdot \boldsymbol{\delta}} + \textrm{B}_{2}^{\textrm{T}} e^{-i\textbf{k}\cdot \boldsymbol{\delta}}\right)\boldsymbol{\hat{\gamma}}_{\textbf{-k}}^{\dagger}\right]
\end{align}
We conclude that the Fourier transform of $\mathcal{\hat{H}}^{(2)}$ can be written as
\begin{equation}\label{H2_SB}
    \mathcal{\hat{H}}^{(2)}_{} = \frac{1}{2} \sum_{\textbf{k}} ( \boldsymbol{\hat{\gamma}}^{\dagger}_{\textbf{k}} \; \boldsymbol{\hat{\gamma}}_{\textbf{-k}}^{})
    \underbrace{\begin{pmatrix}
    A_{\textbf{k}}^{} & B_{\textbf{k}}^{}\\ 
    B_{\textbf{k}}^{*} & A_{\textbf{k}}^{*} \\
\end{pmatrix}}_{\equiv H_{\mathbf{k}}}
\begin{pmatrix}
\boldsymbol{\hat{\gamma}}_{\textbf{k}}^{} \\ \boldsymbol{\hat{\gamma}}^{\dagger}_{\textbf{-k}} \\
\end{pmatrix}
\text{+ const} \ .
\end{equation}
We now diagonalize the $(2d-2)\times (2d-2)$ dimensional matrix
\begin{equation}
    \mathcal{M}_{\textbf{k}}^{} = 
    \begin{pmatrix}
    \unit & 0\\ 
   0 & -\unit \\
\end{pmatrix}
\begin{pmatrix}
    A_{\textbf{k}}^{} & B_{\textbf{k}}^{}\\ 
    B_{\textbf{k}}^{*} & A_{\textbf{k}}^{*} \\
\end{pmatrix} \,,
\end{equation}
where the minus sign is due to the bosonic character of the particles involved. 
\newline 
To conclude, knowing the eigenspace of the matrix $\mathcal{M}_{\textbf{k}}$, the quadratic form in \eqref{H2_SB} can be diagonalized by applying a suitable Bogoliubov rotation of the slave boson operators
\begin{equation}\label{Bogoliubov_rotation}
\begin{split}
    \hat{\gamma}_{\textbf{k},n}^{} &= \sum_{\alpha>0} \left[ u_{\alpha,\textbf{k},n}^{}  \hat{\Gamma}_{\alpha,\textbf{k}}^{} + v^{*}_{\alpha, -\textbf{k},n} \hat{\Gamma}_{\alpha,\textbf{-k}}^\dagger \right]
    \\
    \hat{\gamma}_{\textbf{k},n}^{\dagger} &= \sum_{\alpha>0} \left[ u_{\alpha,\textbf{k},n}^{*} \hat{\Gamma}_{\alpha,\textbf{k}}^{\dagger} + v_{\alpha, -\textbf{k},n}^{} \hat{\Gamma}_{\alpha,\textbf{-k}}^{} \right] \ ,
\end{split}
\end{equation}
where $n$ runs among the levels of excitation.
The diagonalized Hamiltonian then takes the form 
\begin{equation}\label{SB_diagonal_H2}
    \mathcal{\hat{H}}^{(2)} = \sum_{\textbf{k}} \sum_{\alpha>0} \hbar \omega_{\alpha,\textbf{k}}^{} \hat{\Gamma}_{\alpha,\textbf{k}}^{\dagger}\hat{\Gamma}_{\alpha,\textbf{k}}^{} \,,
\end{equation}
from which we get the plot in Fig.~\ref{SM_fullSpectrum}.
\\
\section*{Observables}
Here we show in detail the expressions for the flatness, the density and the loop current shown in the main text. 
We start from the flatness, which is the first order correction to the order parameter. 
From \eqref{btogamma} and thanks to \eqref{Bogoliubov_rotation} it is possible to write the destruction operator as 
\begin{equation}
\begin{split}
    \hat{b}^{}_{i,p} &= [\Tilde{F}_{00}^{i,p}]^* +\\& -  [\Tilde{F}_{00}^{i,p}]^*\sum_{n>0} \frac{1}{N} \sum_{\mathbf{k,k^{\prime}}} e^{-i\mathbf{(k-k^{\prime})\cdot r}_{p}} \sum_{\alpha,\beta>0} \left( u_{\alpha,\mathbf{k},n}^{*} \hat{\Gamma}_{\alpha,\mathbf{k}}^{\dagger} + v_{\alpha, -\mathbf{k},n}^{} \hat{\Gamma}_{\alpha,\mathbf{-k\prime}}^{} \right)\left(u_{\beta,\mathbf{k\prime},n}^{} \hat{\Gamma}_{\beta,\mathbf{k}}^{} + v_{\beta,\mathbf{-k \prime}, n}^{*} \hat{\Gamma}_{\beta,\mathbf{-k\prime}}^{\dagger}\right)+ \\&+ \sum_{n>0} \frac{1}{\sqrt{N}}[\Tilde{F}_{0 n}^{i,p}]^* \sum_{\mathbf{k}} e^{-i\mathbf{k\cdot r}_{p}} \sum_{\alpha>0} \left(u_{\alpha,\mathbf{k},n}^{*} \hat{\Gamma}_{\alpha,\mathbf{k}}^{\dagger} + v_{\alpha, \mathbf{-k},n}^{} \hat{\Gamma}_{\alpha,\mathbf{-k}}^{} \right) + \\&+ \sum_{n>0} \frac{1}{\sqrt{N}} [\Tilde{F}_{n 0}^{i,p}]^* \sum_{\mathbf{k}} e^{i\mathbf{k\cdot r}_{p}} \sum_{\alpha>0} \left( u_{\alpha,\mathbf{k},n}^{} \hat{\Gamma}_{\alpha,\mathbf{k}}^{} + v^{*}_{\alpha, \mathbf{-k},n} \hat{\Gamma}_{\alpha,\mathbf{-k}}^\dagger \right) + \\
    &+\sum_{n,m>0}  [\Tilde{F}_{n m}^{i,p}]^* \frac{1}{N} \sum_{\mathbf{k,k^{\prime}}} e^{-i\mathbf{(k-k^{\prime})\cdot r}_{p}} \sum_{\alpha,\beta>0} \left( u_{\alpha,\mathbf{k},m}^{*} \hat{\Gamma}_{\alpha,\mathbf{k}}^{\dagger} + v_{\alpha,\mathbf{-k}, m}^{} \hat{\Gamma}_{\alpha,\mathbf{-k}}^{} \right)\left(u_{\beta,\mathbf{k\prime},n}^{} \hat{\Gamma}_{\beta,\mathbf{k\prime}}^{} + v_{\beta,\mathbf{-k\prime}, n}^{*} \hat{\Gamma}_{\beta,\mathbf{-k\prime}}^{\dagger}\right) \ . 
\end{split}
\end{equation}
Within linear-response theory, we consider a generic excited state as composed of the ground-state $|\tilde g\rangle$ component and an excitation $|\tilde e \rangle$ component,
\begin{equation}
    |\Tilde{\Psi}_{\alpha}^{}\rangle = |\Tilde{g}\rangle + \varepsilon |\Tilde{e}^{}_{\alpha}\rangle \ ,
\end{equation}
where $\hat{\Gamma}_{\alpha}|\Tilde{g}\rangle = 0$ and $\varepsilon \ll 1$. The expectation value of the destruction operator reads
\be
\mv{\Tilde{\Psi}_{\alpha}|\hat{b}_{i,p}|\Tilde{\Psi}_{\alpha}} = \mv{\tilde{g}|\hat{b}_{i,p}|\tilde{g}} + \varepsilon (\mv{\tilde{g}|\hat{b}_{i,p}|\tilde{e}_{\alpha}} + \mv{\tilde{e}_{\alpha}|\hat{b}_{i,p}|\tilde{g}})=\psi^{i,p} + \varepsilon\, \delta_1\psi^{i,p} +o(\varepsilon^2)\,,
\ee
where the first term is the ground-state superfluid order parameter and the second term its fluctuation due to the excitation $|\tilde e_\alpha\rangle$.
At first order in $\varepsilon$ the only non zero terms are
\be
    \delta_1\hat{\psi}^{i,p} = \frac{1}{\sqrt{N}} \sum_{\alpha>0} \sum_{\textbf{k}} \sum_{n>0} \left[ e^{i\textbf{k}\cdot \textbf{r}_{p}} \left( [\Tilde{F}_{n0}^{i,p}]^*
    u_{\alpha,\textbf{k},n}^{}  + [\Tilde{F}_{0n}^{i,p}]^* v_{\alpha, \textbf{k},n}^{}  \right) \hat{\Gamma}_{\alpha, \textbf{k}}^{} +   e^{-i    
    \textbf{k}\cdot \textbf{r}_{p}} \left( [\Tilde{F}_{0n}^{i,p}]^* u_{\alpha, \textbf{k},n}^{*}  + [\Tilde{F}_{n0}^{i,p}]^* v_{\alpha, \textbf{k},n}^{*}  \right) \hat{\Gamma}_{\alpha, \textbf{k}}^{\dagger}\right] \ ,
\ee
where we can recognize 
\begin{align}
    \mathcal{U}^{i,p}_{\alpha, \textbf{k}} &= \sum_{n>0} \left[  [\Tilde{F}_{n0}^{i,p}]^* u_{\alpha, \textbf{k},n}^{}  + [\Tilde{F}_{0n}^{i,p}]^* v_{\alpha, \textbf{k},n}^{} \right] \\
    \mathcal{V}_{\alpha, \textbf{k}}^{i,p} &= \sum_{n>0} \left[ [\Tilde{F}_{0n}^{i,p}]^* u_{\alpha, \textbf{k},n}^{*}  + [\Tilde{F}_{n0}^{i,p}]^* v_{\alpha, \textbf{k},n}^{*} \right] \,,
\end{align}
that correspond to those used in \cite{DiLibertoMenotti, caleffi2020}.
We can then define the real and imaginary part of $ \delta_1 \psi^{i,p}$ as $\mathcal{R}_{\alpha,\textbf{k}}^{i,p} = |\mathcal{U}_{\alpha,\textbf{k}}^{i,p} + \mathcal{V}_{\alpha,\textbf{k}}^{i,p}|$ and $\mathcal{I}_{\alpha,\textbf{k}}^{i,p} = |\mathcal{U}^{i,p}_{\alpha,\textbf{k}} - \mathcal{V}^{i,p}_{\alpha,\textbf{k}}|$. We are in presence of a pure amplitude mode when $\mathcal{I}^{i,p}_{\alpha,\textbf{k}}=0$, while pure phase excitations of the order parameter correspond to $\mathcal{R}^{i,p}_{\alpha,\textbf{k}}=0$.
\\In this way, the flatness parameter studies the evolution between pure amplitude and phase modes and it is defined as
\begin{equation}
    \mathcal{F}^{i,p}_{\alpha,\textbf{k}} = \frac{\mathcal{R}^{i,p}_{\alpha,\textbf{k}}-\mathcal{I}^{i,p}_{\alpha,\textbf{k}}}{\mathcal{R}^{i,p}_{\alpha,\textbf{k}}+\mathcal{I}^{i,p}_{\alpha,\textbf{k}}} \in \left[ -1,1 \right] \ .
\end{equation} 
In our results, the flatness takes the same value for each site $i$ of the plaquette, thus we will omit the site (and plaquette) indices.
\\Another relevant observable is the density $\hat{n}_{i}^{} = \langle b_{i}^{\dagger} b_{i}^{} \rangle $ for each site $i$. Since this time we are treating $\hat{b}$ operators in the same plaquette (cluster) $p$ we can not consider the expression used in \eqref{bibj_SB_different_plaq}, where we had to consider $\hat{b}$ and $\hat{b}^{\dagger}$ as totally uncorrelated operators.
We derive then $\hat{b}^{\dagger}_{i,p}\hat{b}_{i,p}^{}$ in the following way
\be\label{density_SB_equation}
\begin{split}
    \hat{b}^{\dagger}_{i,p}\hat{b}_{i,p}^{} &= \sum_{n, m}\left( \hat{\gamma}_{n,p}^{\dagger} \tilde{F}_{nm}^{i,p}\hat{\gamma}_{m,p}^{}\right) \times \sum_{n', m'} \left( \hat{\gamma}_{n',p}^{\dagger} [\tilde{F}_{m'n'}^{i,p}]^{*} \hat{\gamma}_{m',p}^{}\right) =\\ &= \sum_{n, m} \left[ \sum_{\mathbf{n}} \left( \mv{\mathbf{n}|\Psi^{(p)}_{n}} \beta_{p,\mathbf{n}}^{\dagger} \right)\mv{\Psi^{(p)}_{n}|\hat{b}_{i,p}^{\dagger}|\Psi^{(p)}_{m}}\sum_{\mathbf{m}} \left( \mv{\Psi^{(p)}_{m}|\mathbf{m}} \beta_{p,\mathbf{m}} \right)\right] \times \\&\times \sum_{n', m'} \left[ \sum_{\mathbf{n'}} \left( \mv{\mathbf{n'}|\Psi_{n'}^{(p)}} \beta_{p,\mathbf{n'}}^{\dagger} \right)\mv{\Psi^{(p)}_{n'}|\hat{b}_{i,p}^{}|\Psi^{(p)}_{m'}}\sum_{\mathbf{m'}} \left( \mv{\Psi^{(p)}_{m'}|\mathbf{m'}} \beta_{p,\mathbf{m'}} \right)\right] = \\&=
    \sum_{\substack{\mathbf{n},\mathbf{n}'\\ \mathbf{m},\mathbf{m}'}} \mv{\mathbf{n}|b_{i,p}^{\dagger}|\mathbf{m}} \mv{\mathbf{n'}|b_{i,p}|\mathbf{m'}} \beta_{p,\mathbf{n}}^{\dagger}\beta_{p,\mathbf{m}}^{}\beta_{p,\mathbf{n'}}^{\dagger}\beta_{p,\mathbf{m'}}^{} = \\ & =      \sum_{\substack{\mathbf{n},\mathbf{n}'\\ \mathbf{m,m'}}} \mv{\mathbf{n}|b_{i,p}^{\dagger}|\mathbf{m}} \mv{\mathbf{n'}|b_{i,p}|\mathbf{m'}} \beta_{p,\mathbf{n}}^{\dagger}\beta_{p,\mathbf{m'}}^{} \delta_{\mathbf{m,n'}} + \cancelto{0}{\sum_{\substack{\mathbf{n},\mathbf{n}'\\ \mathbf{m,m'}}}\mv{\mathbf{n}|b_{i,p}^{\dagger}|\mathbf{m}} \mv{\mathbf{n'}|b_{i,p}|\mathbf{m'}} \beta_{p,\mathbf{n}}^{\dagger}\beta_{p, \mathbf{n'}}^\dagger \beta_{p, \mathbf{m}}^{} \beta_{p,\mathbf{m'}}^{}} = \\ &=  \sum_{\substack{\mathbf{n}\\ \mathbf{m,m'}}} \mv{\mathbf{n}|b_{i,p}^{\dagger}|\mathbf{m}} \mv{\mathbf{m}|b_{i,p}|\mathbf{m'}} \beta_{p,\mathbf{n}}^{\dagger}\beta_{p,\mathbf{m'}}^{} = \\ &=  \sum_{\mathbf{n},\mathbf{m'}} \mv{\mathbf{n}|b_{i,p}^{\dagger}b_{i,p}^{}|\mathbf{m'}} \beta_{p,\mathbf{n}}^{\dagger}\beta_{p,\mathbf{m'}}^{}  \,.
 \end{split}
\ee
The canceled contribution would take nonzero values only when annihilating two occupied states $\mathbf{m}$, $\mathbf{m'}$ via the operators $\beta_{p, \mathbf{m}}^{} \beta_{p,\mathbf{m'}}^{}$, but two occupied SB states are not possible because of the slave boson constraint 
\begin{equation}
    \sum_{\mathbf{n}} \beta_{p,\mathbf{n}}^{\dagger}\beta_{p,\mathbf{n}}^{} = 1\,,
\end{equation}
and we are thus left with the first term only.
Going back to the $\hat{\gamma}$ operators notation we have
\be 
\begin{split}
 \hat{b}^{\dagger}_{i,p}\hat{b}_{i,p} &= \sum_{\mathbf{n},\mathbf{m'}} \sum_{n} \langle \Psi_{n}|\mathbf{n}\rangle \hat{\gamma}_{p,n}^{\dagger}  \mv{\mathbf{n}|b_{i,p}^{\dagger}b_{i,p}^{}|\mathbf{m'}}\sum_{m} \langle \mathbf{m'} |\Psi_{m} \rangle \hat{\gamma}_{p,m}^{} = \\&=
 \sum_{n,m} \mv{\Psi_{n}|b_{i,p}^{\dagger}b_{i,p}^{}|\Psi_{m}} \hat{\gamma}_{p,n}^{\dagger}\hat{\gamma}^{}_{p,m}\,.
 \end{split}
\ee 
In order to perform the condensation on the $\hat{\gamma}_{0}$ operators we expand
\begin{equation}\label{density_SB_full}
\begin{split}
    \hat{n}^{}_{i,p} &= \langle \Psi_{0}|\hat{b}_{i,p}^{\dagger}\hat{b}^{}_{i,p}|\Psi_{0} \rangle \underbrace{\hat{\gamma}_{p,0}^{\dagger}\hat{\gamma}^{}_{p,0}}_{1-\sum_{n>0}\hat{\gamma}_{p,n}^{\dagger}\hat{\gamma}^{}_{p,n}} + \\& + \sum_{n>0} \left[  \hat{\gamma}_{p,n}^{\dagger} \langle \Psi_{n}|\hat{b}_{i,p}^{\dagger}\hat{b}^{}_{i,p}|\Psi_{0} \rangle \hat{\gamma}^{}_{0} +  \hat{\gamma}_{p,0}^{\dagger} \langle \Psi_{0}|\hat{b}_{i,p}^{\dagger}\hat{b}^{}_{i,p}|\Psi_{n} \rangle \hat{\gamma}_{p,n}^{} \right] + \\&+  \sum_{n,m>0}\hat{\gamma}_{p,n}^{\dagger} \langle \Psi_{n}|\hat{b}_{i,p}^{\dagger}\hat{b}^{}_{i,p}|\Psi_{m} \rangle \hat{\gamma}^{}_{p,m}  = \\&=\langle \Psi_{0}|\hat{b}_{i,p}^{\dagger}\hat{b}_{i,p}^{}|\Psi_{0} \rangle \left( 1-\sum_{n>0} \hat{\gamma}_{p,n}^{\dagger} \hat{\gamma}^{}_{p,n} \right) + \sum_{n>0} \left[  \hat{\gamma}_{p,n}^{\dagger} \langle \Psi_{n}|\hat{b}_{i,p}^{\dagger}\hat{b}_{i,p}^{}|\Psi_{0} \rangle  + \langle \Psi_{0}|\hat{b}_{i,p}^{\dagger}\hat{b}_{i,p}^{}|\Psi_{n} \rangle \hat{\gamma}_{p,n}^{} \right] + \\&+  \sum_{n,m>0}\hat{\gamma}_{p,n}^{\dagger} \langle \Psi_{n}|\hat{b}_{i,p}^{\dagger}\hat{b}^{}_{i,p}|\Psi_{m} \rangle \hat{\gamma}^{}_{p,m}  \, .
\end{split}
\end{equation}
The first term in \eqref{density_SB_full} corresponds to the mean-field contribution of the ground-state
\begin{equation}
    \delta_{0} \hat{n}^{}_{i,p} = \langle \Psi_{0}|\hat{b}_{i,p}^{\dagger}\hat{b}_{i,p}^{}|\Psi_{0} \rangle = \sum_{n} |\Tilde{F}^{i,p}_{0n}|^{2} \ .
\end{equation}
The second term is instead the first order correction to the density
\begin{equation}
\delta_{1} \hat{n}^{}_{i,p} = \sum_{n>0,m} \left( \Tilde{F}^{i,p}_{n m}[\Tilde{F}_{0 m}^{i,p}]^*\hat{\gamma}_{p,n}^{\dagger} + \Tilde{F}^{i,p}_{0 m}[\Tilde{F}_{n m}^{i,p}]^* \hat{\gamma}_{p,n} \right) \ ,
\end{equation}
and if we apply the Fourier transform defined before we get
\begin{equation}\label{dens_1stCorr}
\begin{split}
    \delta_{1} \hat{n}^{}_{i,p} &= \frac{1}{\sqrt{N}}\sum_{\alpha > 0}\sum_{\mathbf{k}}\sum_{n>0,m} \left( \Tilde{F}^{i,p}_{nm}[\Tilde{F}_{0m}^{i,p}]^* v_{\alpha, \mathbf{k},n }+ \Tilde{F}_{0m}^{i,p}[\Tilde{F}_{nm}^{i,p}]^* u_{\alpha, \mathbf{k},n} \right) e^{i\mathbf{k \cdot r}_{p}} \hat{\Gamma}_{\alpha,\mathbf{k}} + \\&+ \frac{1}{\sqrt{N}}\sum_{\alpha > 0}\sum_{\mathbf{k}}\sum_{n>0,m} \left( \Tilde{F}^{i,p}_{nm}[\Tilde{F}_{0m}^{i,p}]^* u_{\alpha, \mathbf{k},n}^{*} + \Tilde{F}^{i,p}_{0m}[\Tilde{F}_{nm}^{i,p}]^* v_{\alpha, \mathbf{k},n}^{*} \right) e^{-i\mathbf{k \cdot r}_{p}} \hat{\Gamma}^{\dagger}_{\alpha,\mathbf{k}} \ .
\end{split}
\end{equation}
The second order correction to the density operator is the one quadratic in $\hat{\gamma}_{p,n}$ operators and can be computed from  \eqref{density_SB_full}
\begin{equation} \label{Exc_density_sinSite}
\begin{split}
    \delta_{2} \hat{n}^{}_{i,p} &=
    -\sum_{n} \langle \Psi_{0}|b_{i,p}^{\dagger}|\Psi_{n}\rangle \langle \Psi_{n}|b_{i,p}^{}|\Psi_{0} \rangle\sum_{m>0}\hat{\gamma}_{p,m}^{\dagger}\hat{\gamma}^{}_{p,m} +\sum_{n,m>0}\sum_{l} \hat{\gamma}_{p,n}^{\dagger} \langle \Psi_{n}|b_{i,p}^{\dagger}|\Psi_{l}\rangle \langle \Psi_{l}|b_{i,p}^{}|\Psi_{m} \rangle \hat{\gamma}_{p,m}^{} =\\&= -\sum_{n} |\Tilde{F}^{i,p}_{0n}|^{2}\sum_{m>0}\hat{\gamma}_{p,m}^{\dagger} \hat{\gamma}^{}_{p,m} + \sum_{n,m>0,l} \Tilde{F}^{i,p}_{nl}[\Tilde{F}_{ml}^{i,p}]^* \hat{\gamma}_{p,n}^{\dagger} \hat{\gamma}^{}_{p,m} = \\& =  - \sum_{l} |\Tilde{F}^{i,p}_{0l}|^{2} \sum_{m>0}\frac{1}{N}\sum_{\alpha > 0}\sum_{\mathbf{k},\mathbf{k'}}e^{-i(\mathbf{k-k')\cdot r}_{p}}\hat{\gamma}_{\mathbf{k},m}^{\dagger} \hat{\gamma}_{\mathbf{k'}, m}^{} + \\& + \sum_{n,m>0,l} \Tilde{F}_{nl}^{i,p}[\Tilde{F}_{ml}^{i,p}]^* \frac{1}{N}\sum_{\alpha > 0}\sum_{\mathbf{k},\mathbf{k'}}e^{-i(\mathbf{k-k')\cdot r}_{p}}\hat{\gamma}_{\mathbf{k},n}^{\dagger} \hat{\gamma}_{\mathbf{k'},m}^{} \ .
\end{split}
\end{equation}
From this expression we apply the usual Bogoliubov transformation \eqref{Bogoliubov_rotation}
to get
\begin{equation}
\begin{split}
    \delta_{2} \hat{n}_{i,p}^{} &= -\sum_{l} |\Tilde{F}^{i,p}_{0l}|^{2} \sum_{m>0}\frac{1}{N}\sum_{\mathbf{k},\mathbf{k'}}e^{-i(\mathbf{k-k')\cdot r}_{p}}\sum_{\alpha,\beta>0} \left( u_{\alpha, \mathbf{k},m}^{*} \hat{\Gamma}^{\dagger}_{\alpha,\mathbf{k}} + v^{}_{\alpha, \mathbf{-k},m} \hat{\Gamma}_{\alpha,\mathbf{-k}}^{} \right)\left( u_{\beta, \mathbf{k'},m}^{} \hat{\Gamma}_{\beta,\mathbf{k'}}^{} + v_{\beta ,\mathbf{-k'},m}^{*} \hat{\Gamma}_{\beta,\mathbf{-k'}}^{\dagger} \right) + \\&
    + \sum_{n,m>0,l} \Tilde{F}_{nl}^{i,p}[\Tilde{F}_{ml}^{i,p}]^* \frac{1}{N}\sum_{\mathbf{k},\mathbf{k'}}e^{-i(\mathbf{k-k')\cdot r}_{p}}\sum_{\alpha,\beta>0} \left( u_{\alpha ,\mathbf{k},n}^{*} \hat{\Gamma}^{\dagger}_{\alpha,\mathbf{k}} + v_{\alpha ,\mathbf{-k},n}^{} \hat{\Gamma}_{\alpha,\mathbf{-k}}^{} \right)\left( u_{\beta, \mathbf{k'},m}^{} \hat{\Gamma}_{\beta,\mathbf{k'}}^{} + v_{\beta ,\mathbf{-k'},m}^{*} \hat{\Gamma}_{\beta,\mathbf{-k'}}^{\dagger} \right) \ .
\end{split}
\end{equation}
\\ With \eqref{Exc_density_sinSite} we can also compute the expectation value of the density on the excited states. 
Indicating
$|\Bar{\alpha},\mathbf{\Bar{k}}\rangle$ [$\equiv \hat{\Gamma}^\dagger_{\bar \alpha,\bar{\mathbf k}}|\tilde g\rangle$] as the excited state, the density expectation value reads
\begin{equation}
    \langle \hat{n}_{i,p}^{} \rangle_{\Bar{\alpha}^{},\mathbf{\Bar{k}}} = \mv{\delta_{0}\hat{n}_{i,p}} + \langle \delta_{2} \hat{n}^{}_{i,p} \rangle_{\Bar{\alpha},\mathbf{\Bar{k}}} \ .
\end{equation} 
Here $n^{0}_{i,p}$ is the mean-field density of the ground-state for the site $i$ of the $p$-th plaquette. In particular we get
\begin{equation}\label{dens_excState}
\begin{split}
    \langle \delta_{2} \hat{n}^{}_{i,p} \rangle_{\Bar{\alpha},\mathbf{\Bar{k}}} =& -\frac{1}{N}\sum_{l}\left[|\Tilde{F}^{i,p}_{0l}|^{2}\sum_{m>0}\left(|u_{\Bar{\alpha},\mathbf{\Bar{k}},m}^{}|^{2} + |v_{\Bar{\alpha},\mathbf{\Bar{k}},m}^{}|^{2} + \sum_{\mathbf{k}} \sum_{\alpha>0} |v_{\alpha,\mathbf{k},m}^{}|^{2}\right)\right] + \\ & +\frac{1}{N}\sum_{n,m>0,l} \left[ \Tilde{F}^{i,p}_{nl}[\Tilde{F}_{ml}^{i,p}]^* \left(u_{\Bar{\alpha},\mathbf{\Bar{k}},n}^{*}u_{\Bar{\alpha},\mathbf{\Bar{k}},m}^{} + v_{\Bar{\alpha},\mathbf{\Bar{k}},n}^{*}v_{\Bar{\alpha},\mathbf{\Bar{k}},m}^{} +   \sum_{\alpha>0}\sum_{\mathbf{k}} 
    v_{\alpha,\mathbf{k},n}^{*}v_{\alpha,\mathbf{k},m}^{} \right)\right] = \\&= -\frac{1}{N}\sum_{l}\left[|\Tilde{F}^{i,p}_{0l}|^{2}\sum_{m>0}\left(|u_{\Bar{\alpha},\mathbf{\Bar{k}},m}^{}|^{2} + |v_{\Bar{\alpha},\mathbf{\Bar{k}},m}^{}|^{2}\right)\right]+\frac{1}{N}\sum_{n,m>0,l} \left[ \Tilde{F}^{i,p}_{nl}[\Tilde{F}_{ml}^{i,p}]^* \left(u_{\Bar{\alpha},\mathbf{\Bar{k}},n}^{*}u_{\Bar{\alpha},\mathbf{\Bar{k}},m}^{} + v_{\Bar{\alpha},\mathbf{\Bar{k}},n}^{*}v_{\Bar{\alpha},\mathbf{\Bar{k}},m}^{}    \right)\right] +\\ & + \sum_\mathbf{k}\left( \dots \right)\,.
\end{split}
\end{equation}
From this expression, we derive the results for the density variations of the different modes displayed in the main text in the saddle point approximation, namely we did not include the quantum fluctuations (i.e., the $\sum_{\mathbf{k}}$ terms in the previous expression).
Notice that for the Goldstone mode ($\bar \alpha = 1$), a divergence occurs when including for $\bar{\mathbf k} =0$ since the Bogoliubov coefficients $v_{1,\mathbf{k},n}$ diverge at $\mathbf{k}\rightarrow 0$, as in standard bosonic gases.
\\ The last observable we want to compute is the loop current. We focus on the correlator $\langle \hat{b}_{i,p}^{\dagger} \hat{b}_{j,p} \rangle$: the loop current will be a linear combination of some of these terms for each plaquette, as shown in the main text. 
We can write 
\begin{equation}
    \langle \hat{b}_{i,p}^{\dagger}\hat{b}_{j,p}\rangle_{\Bar{\alpha},\mathbf{\Bar{k}}} = \mv{\delta_{0} (\hat{b}_{i,p}^{\dagger}\hat{b}_{j,p}^{})} + \mv{\delta_{1} (\hat{b}_{i,p}^{\dagger}\hat{b}_{j,p}^{})} + \mv{\delta_{2} (\hat{b}_{i,p}^{\dagger}\hat{b}_{j,p}^{})} \,,
\end{equation}
The ground-state value is 
\begin{equation}
     \langle \delta_{0} (\hat{b}_{i,p}^{\dagger}\hat{b}_{j,p}^{}) \rangle  =\sum_{n} \Tilde{F}_{0n}^{i,p}[\Tilde{F}_{0n}^{j,p}]^* \ .
\end{equation}
Recalling \eqref{dens_1stCorr} it is possible to show the first order correction to this two site correlator, which takes the form 
\begin{equation}
\begin{split}
    \delta_{1} (\hat{b}_{i,p}^{\dagger}\hat{b}_{j,p}^{}) &=  \frac{1}{\sqrt{N}}\sum_{\mathbf{k}}\sum_{n>0,m} \sum_{\alpha>0} \left( \Tilde{F}^{i,p}_{nm}[\Tilde{F}^{j,p}_{0m}]^* v_{\alpha,\mathbf{k},n}^{} + \Tilde{F}^{i,p}_{0m}[\Tilde{F}^{j,p}_{nm}]^* u_{\alpha, \mathbf{k},n}^{} \right) e^{i\mathbf{k \cdot r}_{p}} \hat{\Gamma}_{\alpha,\mathbf{k}}^{} + \\&+  \frac{1}{\sqrt{N}}\sum_{\mathbf{k}}\sum_{n>0,m} \sum_{\alpha>0} \left( \Tilde{F}^{i,p}_{nm}[\Tilde{F}^{j,p}_{0m}]^* u_{\alpha, \mathbf{k},n}^{*} + \Tilde{F}^{i,p}_{0m}[\Tilde{F}^{j,p}_{nm}]^* v_{\alpha, \mathbf{k},n}^{*} \right) e^{-i\mathbf{k \cdot r}_{p}} \hat{\Gamma}^{\dagger}_{\alpha,\mathbf{k}} \ .
\end{split}
\end{equation}
The second order correction can be instead written as
\begin{equation}
\begin{split}
    \delta_{2} (\hat{b}_{i,p}^{\dagger}\hat{b}_{j,p}^{}) &= -\sum_{l} \Tilde{F}^{i,p}_{0l}[\Tilde{F}^{j,p}_{0l}]^* \sum_{m>0}\frac{1}{N}\sum_{\mathbf{k},\mathbf{k'}}e^{-i(\mathbf{k-k')\cdot r}_{p}}\sum_{\alpha,\beta>0} \left( u_{\alpha, \mathbf{k},m}^{*} \hat{\Gamma}^{\dagger}_{\alpha,\mathbf{k}} + v_{\alpha, \mathbf{-k},m}^{} \hat{\Gamma}_{\alpha,\mathbf{-k}}^{} \right)\left( u_{\beta, \mathbf{k'},m}^{} \hat{\Gamma}_{\beta,\mathbf{k'}}^{} + v_{\alpha ,\mathbf{-k'},m}^{*} \hat{\Gamma}_{\beta,\mathbf{-k'}}^{\dagger} \right) + \\&
    + \sum_{n,m>0,l} \Tilde{F}_{nl}^{i,p}[\Tilde{F}_{ml}^{j,p}]^* \frac{1}{N}\sum_{\mathbf{k},\mathbf{k'}}e^{-i(\mathbf{k-k')\cdot r}_{p}}\sum_{\alpha,\beta>0} \left( u_{\alpha ,\mathbf{k},n}^{*} \hat{\Gamma}^{\dagger}_{\alpha,\mathbf{k}} + v_{\alpha ,\mathbf{-k},n}^{} \hat{\Gamma}_{\alpha,\mathbf{-k}}^{} \right)\left( u_{\beta, \mathbf{k'},m}^{} \hat{\Gamma}_{\beta,\mathbf{k'}}^{} + v_{\alpha ,\mathbf{-k'},m}^{*} \hat{\Gamma}_{\beta,\mathbf{-k'}}^{\dagger} \right) \ .
\end{split}
\end{equation}
\begin{figure}
    \centering  \includegraphics[width=0.98\linewidth]{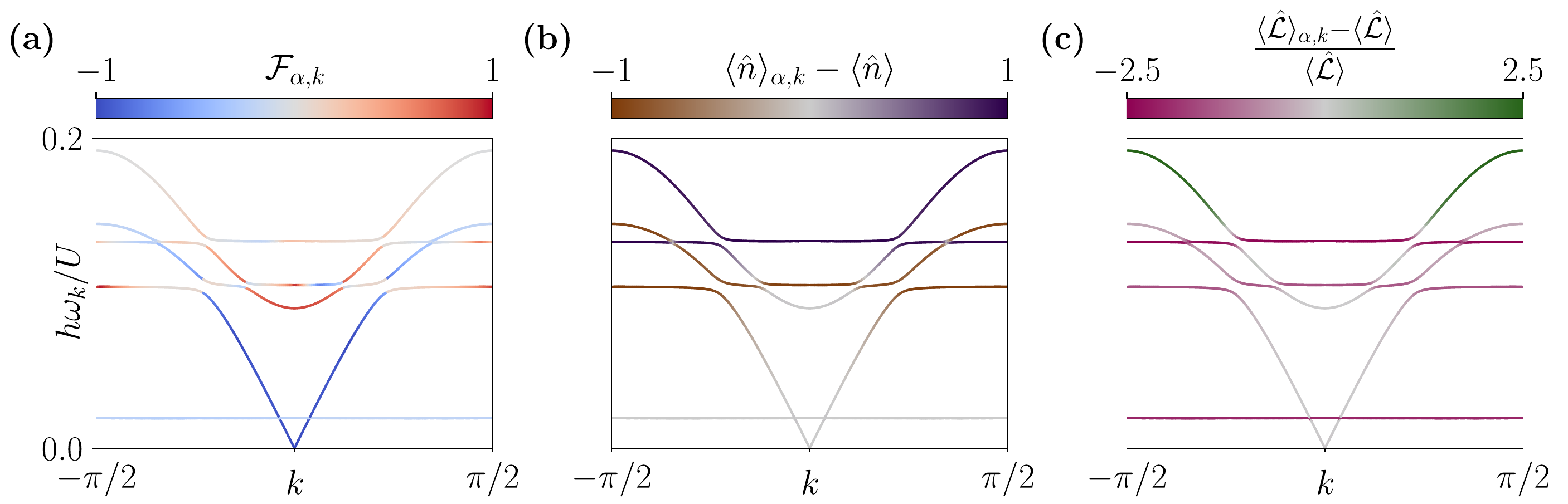}
    \caption{Low-energy excitation spectrum for filling $\nu = 1/2$ at $J'/U = 0.02$, $J/U = 0.67$, and $\mu/U = -0.67$, with $|\mathbf{k}_x| = |\mathbf{k}_y| = k$.
\textbf{(a)} Flatness $\mathcal{F}_{\alpha,k}$ of the excitation bands, showing the gapless Goldstone mode (blue) and an amplitude (Higgs) mode (red). The flat bands present different values of flatness due to band crossings.
\textbf{(b)} Plaquette density, showing two bands with particle and hole excitations.
\textbf{(c)} Loop current $\langle \mathcal{\hat{L}}_{\alpha,k} \rangle$ for each band relative to the ground-state current $\langle \mathcal{\hat{L}} \rangle$. The Higgs mode carries a loop current with the same sign as the ground-state one.}
    \label{nu=1/2}
\end{figure}
In order to compute the loop current for the bands in the excitation spectra, since $\mv{    \delta_{1} (\hat{b}_{i,p}^{\dagger}\hat{b}_{j,p}^{}) }=0$ we can write
\begin{equation}
    \langle \hat{b}_{i,p}^{\dagger}\hat{b}_{j,p}^{} \rangle^{}_{\Bar{\alpha},\mathbf{\Bar{k}}} = \mv{\delta_{0} (\hat{b}_{i,p}^{\dagger}\hat{b}_{j,p}^{})} + \langle \delta_{2} (\hat{b}_{i,p}^{\dagger}\hat{b}_{j,p}^{} )\rangle^{}_{\Bar{\alpha},\mathbf{\Bar{k}}}
\end{equation}
If we focus on the last term we can get
\begin{equation}
\begin{split}
    \mv{\delta_{2} (\hat{b}_{i,p}^{\dagger}\hat{b}_{j,p}^{})}_{\Bar{\alpha},\mathbf{\Bar{k}}} &= -\sum_{n} \Tilde{F}^{i,p}_{n0}[\Tilde{F}^{j,p}_{n0}]^* \sum_{m>0} \left( |u_{\Bar{\alpha},\Bar{\mathbf{k}},m}^{}|^{2} + |v_{\Bar{\alpha},\Bar{\mathbf{k}},m}^{}|^{2} + \frac{1}{N}\sum_{\mathbf{k}}\sum_{\alpha>0} |v_{\alpha,\mathbf{k},m}^{}|^{2}\right) +\\
    &+ \sum_{n,m>0} \sum_{l}\left[ \Tilde{F}_{nl}^{i,p}[\Tilde{F}_{ml}^{j,p}]^* \left( u_{\Bar{\alpha},\Bar{\mathbf{k}},n}^{*}u_{\Bar{\alpha},\Bar{\mathbf{k}},m}^{} + v_{\Bar{\alpha},\Bar{\mathbf{k}},n}^{}v_{\Bar{\alpha},\Bar{\mathbf{k}},m}^{*} + \frac{1}{N}\sum_{\mathbf{k}}\sum_{\alpha>0} v_{\alpha,\mathbf{k},n}^{} v_{\alpha,\mathbf{k},m}^{*}  \right) \right] = \\&= -\sum_{n} \Tilde{F}^{i,p}_{n0}[\Tilde{F}^{j,p}_{n0}]^* \sum_{m>0} \left( |u_{\Bar{\alpha},\Bar{\mathbf{k}},m}^{}|^{2} + |v_{\Bar{\alpha},\Bar{\mathbf{k}},m}^{}|^{2}  \right) + \sum_{n,m>0} \sum_{l}\left[ \Tilde{F}_{nl}^{i,p}[\Tilde{F}_{ml}^{j,p}]^* \left( u_{\Bar{\alpha},\Bar{\mathbf{k}},n}^{*}u_{\Bar{\alpha},\Bar{\mathbf{k}},m}^{} + v_{\Bar{\alpha},\Bar{\mathbf{k}},n}^{}v_{\Bar{\alpha},\Bar{\mathbf{k}},m}^{*}  \right) \right] \\ & + \sum_\mathbf{k} \left( \dots \right) \ ,
\end{split}
\end{equation}
which is the expression we used to compute the loop current for the lowest excitation bands. As for the density, we neglect the quantum fluctuations and the $\mathbf{\bar{k}=0}$ point for $\bar{\alpha}=1$ (Goldstone). In the main text, we focused on the transition between the trivial Mott insulating and vortex superfluid phases at integer filling. Here, we extend the analysis to fractional fillings using the Gutzwiller mean-field approach. For the three cases of interest, $\nu = 3/4$, $1/2$, and $1/4$, we find the same qualitative behavior: a transition between a vortex insulating and a vortex superfluid phase.
As shown in Fig.~\ref{nu=1/2}, which reports the excitation spectrum for $\nu = 1/2$, the low-energy physics is governed by five excitation bands. Panel (a) reveals the presence of a gapless Goldstone mode and a single Higgs (amplitude) mode. In addition, we identify particle and hole branches, together with a fifth flat band characterized by nearly constant density and a flatness $\mathcal{F}_{k\sim \pi/100} \approx 0$.
From panel (c), we infer that this flat band carries an angular momentum opposite to that of the ground-state. In contrast, the Higgs mode exhibits the same angular momentum as the ground-state $\langle \hat{\mathcal{L}} \rangle$. This amplitude mode hybridizes with the Goldstone mode at the transition point, giving rise to a Vortex-Mott Insulating phase. In this case, only the U(1) symmetry is broken, since the discrete $\mathbb{Z}_{2}$ symmetry is broken also in the vortex-superfluid phase. 
In such time-reversal broken phase, the low-energy behaviour described by the quasiparticles is not different from the one in the conventional Bose-Hubbard model. As we have seen, a different scenario instead takes place at integer filling.
\section*{Additional DMRG results and details.}
To further support the cluster Gutzwiller results, we perform density-matrix renormalization group (DMRG) simulations of the bosonic BBH model on a ladder geometry. We analyze several observables to characterize the phase transitions at fillings $\nu=1$ and $\nu=1/4$. A central quantity is the fidelity susceptibility,
\begin{equation}
\label{chi_eq}
\chi_{\mathcal{F}} = -\frac{1}{L}
\frac{\langle \psi(U+\delta U) | \psi(U) \rangle  - 2 + \langle \psi(U-\delta U) | \psi(U) \rangle}{\delta U^{2}} \,,
\end{equation}
where $L=2 \times L_x$ is the total number of sites and $|\psi(U)\rangle$ denotes the ground-state matrix-product state.

For filling $\nu=1$, the results are presented in the main text. Convergence was achieved using a maximum bond dimension $\chi_{\mathrm{max}}=1800$ for $L_x=40,44,48$, and $\chi_{\mathrm{max}}=1600$ for $L_x=32$. The singular-value decomposition cutoff was set to $10^{-10}$ for all system sizes, yielding an energy convergence better than $5\times10^{-7}$ in all cases. The displacement in the interaction strength in Eq.~(\ref{chi_eq}) was fixed to $\delta U=0.002$.

We were unable to reliably identify the $\nu=3/4$ transition within the present DMRG simulations, and a detailed study of this filling is left for future work, while the case $\nu=1/2$ was already investigated in Ref.~\cite{DiLibertoGoldman}.

\begin{figure}
    \centering    \includegraphics[width=
\linewidth]{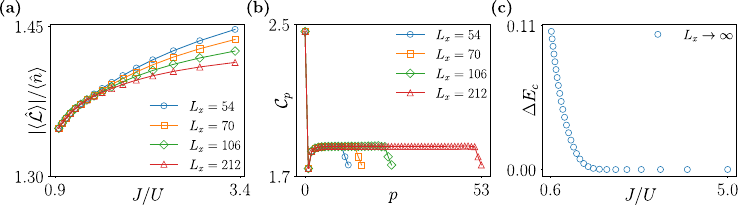}
    \caption{Additional DMRG results for $\nu=1/4$ for $J'=0.1J$. \textbf{(a)}: $|\langle \hat{\mathcal{L}} \rangle| / \langle \hat{n} \rangle $ for different ladder lengths $L=54,70,106,212$. \textbf{(b)}: Loop current-loop current correlation throughout all the transition. \textbf{(c)}: Charge gap $\Delta E_{c}$ computed as the fit to $L \to \infty$ from the value of $L_{x}$=70, 98, 106.}
    \label{DMRG_nu0.25}
\end{figure}

The $\nu=1/4$ transition is discussed in the main text, while additional numerical results are presented in Fig.~\ref{DMRG_nu0.25}. Figs.~\ref{DMRG_nu0.25}(a) and (b) show the behavior of $\langle \hat{\mathcal{L}}\rangle $ and the plaquette 
loop current correlator $\mathcal{C}_{p}$ for different system lengths. Figure~\ref{DMRG_nu0.25}(c) displays the opening of the charge gap, defined as $\Delta E_{c} = E(N+1) - 2E(N) + E(N-1)$. The nature of the transition can be inferred from panel (c): the emergence of a finite $\Delta E_{c}$ indicates that the system becomes incompressible, signaling the onset of a Mott insulating phase. Concurrently, the finite loop current confirms the chiral character of the superfluid phase.

The effective theory developed in Ref.~\cite{DiLibertoGoldman} shows that the low-energy physics of weakly interacting bosons in a $\pi$-flux plaquette shares important similarities with that of $p$-band bosons.
However, crucial differences occur, one example is the transition at filling $\nu=1/4$. 
Indeed, in $p$-band models the strong anisotropy of orbital hopping induces a transition into a non-chiral phase described by a classical product state theory \cite{li2016physics, Li_TRS}. 
Here the situation changes as the emergent orbitals hop with the same strength in all spatial directions ($J'$). 
The Gutzwiller ansatz predicts a transition into a vortex-chiral Mott phase. 
To further substantiate the nature of this vortex phase, we employ a perturbative approach analogous to that used in Refs.~\cite{li2016physics, Li_TRS}. 
As stated in the main text, at second order in perturbation theory with $\hat \sigma^z_p$ corresponding to loop current states, one obtains
\be\label{Pert_ham_nu1_over_4}
\hat H_{\text{eff}}= - \frac{5}{2}\frac{J'^{2}}{U} -\frac{J'^{2}}{U}\sum_{\mv{p,q}} \Bigl[
\frac{3}{4}\hat{\sigma}^{z}_p\hat{\sigma}^{z}_q
+ \frac{1}{4}\bigl(\hat{\sigma}^{x}_p\hat{\sigma}^{x}_q
+ \hat{\sigma}^{y}_p\hat{\sigma}^{y}_q\bigr) \Bigr] 
\ee
which suggests a chiral ferromagnetic ground-state in this regime, explaining the chiral gapped order.
Notice that differently from $p$-bands, not only the phase is chiral but also it is described at leading order by a quantum theory and not a classical (Ising) model.

\section*{Flux-quench protocol}

\begin{figure}[h!]
    \centering

    \includegraphics[width=0.3\linewidth]{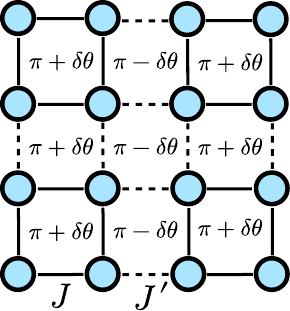}
    \caption{Initial staggered flux configuration used for quenching the system in the superfluid ordered phase.}
    \label{flux_staggered}
\end{figure}

To excite chiral collective modes, we implement a quench that modulates the synthetic magnetic flux pattern across the lattice. Specifically, we introduce a staggered flux perturbation along the $x$ direction such that neighboring plaquettes acquire alternating flux values, as illustrated in Fig.~\ref{flux_staggered}.
The modulation is implemented by adjusting the Peierls phases on the hoppings, while keeping their magnitudes unchanged. This perturbation explicitly breaks time-reversal symmetry while preserving the global U(1) particle-number symmetry. As a result, the quench predominantly couples to current-carrying excitations rather than density fluctuations.

Since our goal is to probe amplitude excitations such as the chiral Higgs mode (H1), we apply the perturbation only within the superfluid phase. The quench amplitude is chosen to be small ($\delta\theta \ll 1$) in order to remain in the linear-response regime. Concretely, we set
\begin{equation}
\delta\theta =
0.01 \left(\dfrac{J-\bar{J}}{\bar{J}}\right), \qquad \text{with} \quad J > \bar{J}
\end{equation}
where $\bar{J}/U = 0.88$ is chosen such that the cluster Gutzwiller ground-state lies in the superfluid regime.

After quenching the system with $\delta \theta \to 0$, we monitor the real-time evolution of the plaquette loop current and the local superfluid order parameter by solving the time evolution within the cGWA approach. The Fourier analysis of the resulting oscillations and the corresponding power spectrum $P(\omega)$ reveals a dominant spectral peak at the frequency of the lowest chiral Higgs mode, $\hbar\omega_{\rm{H1}}$, as discussed in the main text, demonstrating that the staggered-flux protocol provides a probe of TRS–odd amplitude excitations.
Notice that a very weak signal at frequency 2$\hbar\omega_{\rm{H1}}$, which we interpret as a higher-harmonic generation mechanism due to the intrinsic nonlinearity of the cGWA. The quench also displays the excitation of H2. However all these signature are several orders of magnitudes smaller than the dominant H1 signal.

\section*{Additional TTN analysis}
The DMRG analysis shown in the main text in the previous sections focused on ladder geometries where we fixed $L_y=2$. 
To go beyond that and investigate similar physics in a more extended two-dimensional geometry, we employed Tree Tensor Network (TTN) ansatz \cite{TTN_Tagliacozzo,FelserMontangero2026_TTN,TTN_BH}. 
Since a TTN maps the lattice onto a linear network via a space-filling curve (here we used a snake mapping), even a local 2D model acquires effective long-range interactions, which the network encodes without loss of efficiency: TTNs remain loopless, and thus efficiently contractible, with all MPS algorithms that carry over directly.

Using this approach, we simulated an $(L_x,L_y) = (8,8)$ system with periodic boundary conditions and an average density of one boson per site. The state is represented as a binary tree of tensors, with physical degrees of freedom on the bottom layer and auxiliary tensors above transferring long-range correlations across the network.
\\The simulations were carried out using the Quantum Tea library~\cite{qtealeaves2026,qredtea,qredteaArticle2024}. The maximum bond dimension used in all simulation is $\chi=400$, while the local Hilbert space is truncated at occupation $n_{\mathrm{max}}=2$. All simulations were converged to a relative energy tolerance $\Delta E_{\mathrm{rel}}<10^{-5}$, where $\Delta E_{\mathrm{rel}} = \textrm{max}(|E-E_{\textrm{last}}|)/(4|E_{\textrm{last}}|)$ is evaluated over the last four recorded energy values, with $E_{\textrm{last}}$ denoting the most recent energy estimate. To avoid the convergence to the superposition of the two ground-state with opposite chirality we explicitly break $\mathbb{Z}_2^T$ by introducing a small $\delta \phi = 10^{-3}$ in $y=1,3,5,7$ legs of the lattice.

\begin{figure}
    \centering
    \includegraphics[width=0.98\linewidth]{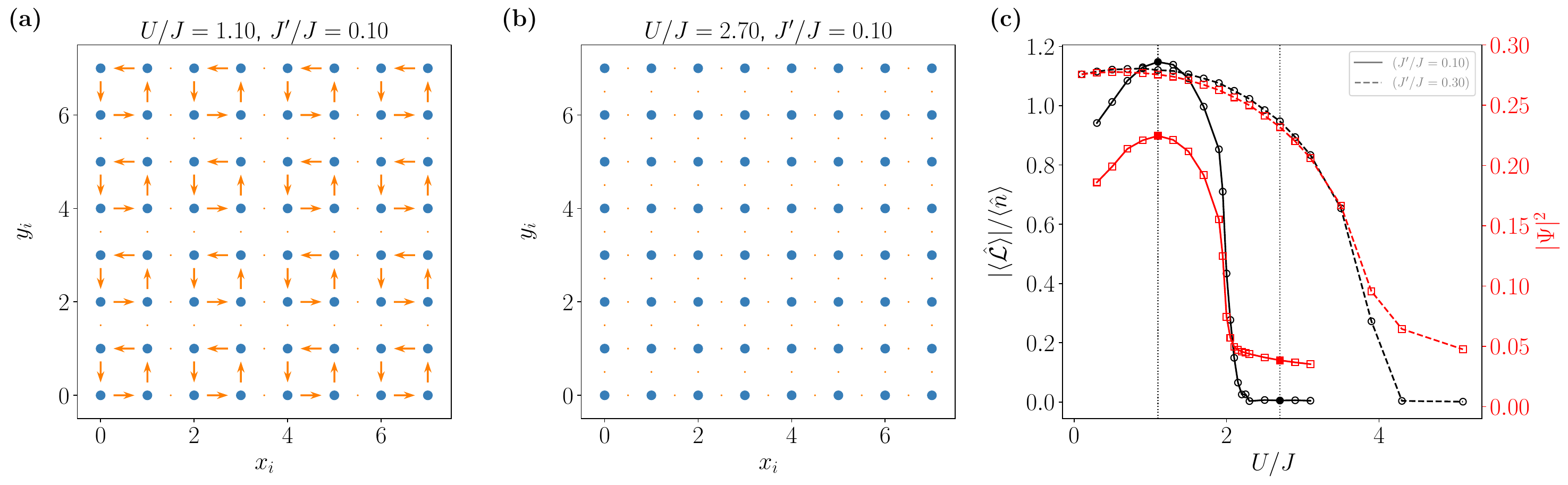}
    \caption{TTN results for an $8\times8$ lattice with periodic boundary conditions in both directions at fixed filling $\nu=1$, with $J'/J=0.10$ \textbf{(a,b,c)} and $J'/J=0.30$ \textbf{(c)}. The local Hilbert space is truncated at $n_{\mathrm{max}}=2$, the maximum bond dimension is $\chi=400$. All simulations were converged to an absolute energy tolerance $\Delta E_{\mathrm{abs}}=|E(n)-E(n-1)|<10^{-7}$, where $n$ denotes the sweep number. \textbf{(a,b)} Bond-current expectation values, defined in the main text, for $U/J=1.10$ and $U/J=2.70$, respectively. The largest bond currents are $\langle \hat{j}_{(4,1)\rightarrow(5,1)}\rangle=1.149 J$ in panel \textbf{(a)} and $\langle \hat{j}_{(4,3)\rightarrow(5,3)}\rangle=0.007 J$ in panel \textbf{(b)}. 
    \textbf{(c)} Loop current $\langle\hat{\mathcal{L}}\rangle$ (black circles) on the plaquette $(2,2)\rightarrow(3,2)\rightarrow(3,3)\rightarrow(2,3)\rightarrow(2,2)$ and condensate fraction $|\Psi|^2$ (red squares) as functions of $U/J$ for $J'/J=0.10$. The filled symbols indicate the interaction strengths corresponding to panels \textbf{(a)} and \textbf{(b)}. Both observables vanish at approximately $U/J\sim 2.20$. The condensate fraction and loop current behaviours are also shown for $J'/J=0.30$ (dashed lines), where $\mv{\hat{\mathcal{L}}}$ (black squares) and $|\Psi|^2$ (red diamonds) drop at $U/J \sim 4.10$.}
    \label{fig:SM6}
\end{figure}

In Figs.~\ref{fig:SM6}a and \ref{fig:SM6}b we show a vortex-array regime ($U/J=1.10$) and a vortex-array free one ($U/J=2.70$), respectively, for $J'/J=0.1$. As shown in Fig.~\ref{fig:SM6}c, the loop current (black circles) and the condensate fraction (red triangles),
\[
|\Psi|^2=\frac{1}{L^2}\sum_{ij}\langle \hat{b}_i^\dagger\hat{b}_j\rangle,
\]
both decrease sharply around $U/J\sim2.20$ $J'/J=0.10$ (solid lines). 
For comparison, we also report the corresponding quantities for $J'/J=0.30$ (dashed lines), where the condensate fraction (red diamonds) and the loop current (black squares) drop for larger values of $U/J$. 
In both cases, the two orders appear to be tied together in a broader range of parameters.
Within the finite and approximate method used the two observables show a maximum drop rate around the same interaction strength.
However, these finite-size calculations alone do not allow us to determine whether the restoration of the U(1) and $\mathbb{Z}_2^T$ symmetries occurs through a single phase transition or through two distinct but closely spaced transitions.

\end{document}